\documentclass{article}

\usepackage{arxivsimplified}

\usepackage{moreverb}
\usepackage{cite}
\usepackage{amsmath,amsfonts,amssymb}
\usepackage{subfigure}
\usepackage{float}
\usepackage{color}
\usepackage{multirow}
\usepackage{graphicx}
\graphicspath{{Figures/}}
\usepackage{mathtools}
\usepackage{enumitem}

\usepackage{booktabs}

\usepackage[maxfloats=100]{morefloats}
\usepackage{relsize}

\usepackage[colorlinks,bookmarksopen,bookmarksnumbered,citecolor=red,urlcolor=red]{hyperref}

\newcommand\BibTeX{{\rmfamily B\kern-.05em \textsc{i\kern-.025em b}\kern-.08em
T\kern-.1667em\lower.7ex\hbox{E}\kern-.125emX}}

\usepackage[utf8]{inputenc} 
\usepackage[T1]{fontenc}    
\usepackage{hyperref}       
\usepackage{url}            
\usepackage{booktabs}       
\usepackage{amsfonts}       
\usepackage{nicefrac}       
\usepackage{microtype}      
\usepackage{lipsum}

\usepackage{graphicx}
\usepackage{cite}
\usepackage[dvipsnames]{xcolor}
\usepackage{amsmath}
\usepackage{blindtext}
\usepackage{amsfonts}
\usepackage{multirow}
\usepackage{array}
\usepackage{mathtools}
\usepackage{listings}
\usepackage{xcolor}
\definecolor{codegreen}{rgb}{0,0.6,0}
\definecolor{codegray}{rgb}{0.5,0.5,0.5}
\definecolor{codepurple}{rgb}{0.58,0,0.82}
\definecolor{backcolour}{rgb}{0.95,0.95,0.92}

\lstdefinestyle{mystyle}{
    backgroundcolor=\color{backcolour},   
    commentstyle=\color{codegreen},
    keywordstyle=\color{magenta},
    numberstyle=\tiny\color{codegray},
    stringstyle=\color{codepurple},
    basicstyle=\ttfamily\footnotesize,
    breakatwhitespace=false,         
    breaklines=true,                 
    captionpos=b,                    
    keepspaces=true,                 
    numbers=left,                    
    numbersep=5pt,                  
    showspaces=false,                
    showstringspaces=false,
    showtabs=false,                  
    tabsize=2
}

\usepackage{graphics}
\usepackage{adjustbox}
\usepackage{algorithm}
\usepackage[noend]{algpseudocode}

\makeatletter
\newenvironment{breakablealgorithm}
  {
   \begin{center}
     \refstepcounter{algorithm}
     \hrule height.8pt depth0pt \kern2pt
     \renewcommand{\caption}[2][\relax]{
       {\raggedright\textbf{\ALG@name~\thealgorithm} ##2\par}%
       \ifx\relax##1\relax 
         \addcontentsline{loa}{algorithm}{\protect\numberline{\thealgorithm}##2}%
       \else 
         \addcontentsline{loa}{algorithm}{\protect\numberline{\thealgorithm}##1}%
       \fi
       \kern2pt\hrule\kern2pt
     }
  }{
     \kern2pt\hrule\relax
   \end{center}
  }
\makeatother

\usepackage{scalerel,stackengine}
\stackMath
\newcommand\reallywidehat[1]{%
\savestack{\tmpbox}{\stretchto{%
  \scaleto{%
    \scalerel*[\widthof{\ensuremath{#1}}]{\kern-.6pt\bigwedge\kern-.6pt}%
    {\rule[-\textheight/2]{1ex}{\textheight}}
  }{\textheight}%
}{0.5ex}}%
\stackon[1pt]{#1}{\tmpbox}%
}

\title{A long short-term memory embedding for hybrid uplifted reduced order models}

\author{
  Shady~E.~Ahmed \\
  School of Mechanical \& Aerospace Engineering,\\
  Oklahoma State University, \\
  Stillwater, Oklahoma - 74078, USA.\\
  \texttt{shady.ahmed@okstate.edu} \\
  \And
  Omer San \\
  School of Mechanical \& Aerospace Engineering,\\
  Oklahoma State University, \\
  Stillwater, Oklahoma - 74078, USA.\\
  \texttt{osan@okstate.edu} \\
  \And
 Adil Rasheed \qquad \qquad \quad \\
 Department of Engineering Cybernetics, \qquad \qquad \qquad\\
  Norwegian University of Science and Technology, \qquad \qquad \qquad \\
  N-7465, Trondheim, Norway. \qquad \qquad \quad\\
  \texttt{adil.rasheed@ntnu.no \qquad \qquad  \qquad}\\
  \And
  Traian Iliescu \qquad \qquad \quad \\
  Department of Mathematics, \qquad \qquad \qquad \\
  Virginia Tech, \qquad \qquad \qquad \\
  Blacksburg, VA 24061, USA. \qquad \qquad \qquad\\
  \texttt{iliescu@vt.edu \qquad \quad  \quad}\\
}

\begin{document}
\maketitle

\begin{abstract}
In this paper, we introduce an uplifted reduced order modeling (UROM) approach through the integration of standard projection based methods with long short-term memory (LSTM) embedding. Our approach has three modeling layers or components. In the first layer, we utilize an intrusive projection approach to model dynamics represented by the largest modes. The second layer consists of an LSTM model to account for residuals beyond this truncation. This closure layer refers to the process of including the residual effect of the discarded modes into the dynamics of the largest scales. However, the feasibility of generating a low rank approximation tails off for higher Kolmogorov $n$-width systems due to the underlying nonlinear processes. The third uplifting layer, called super-resolution, addresses this limited representation issue by expanding the span into a larger number of modes utilizing the versatility of LSTM. Therefore, our model integrates a physics-based projection model with a memory embedded LSTM closure and an LSTM based super-resolution model. In several applications, we exploit the use of Grassmann manifold to construct UROM for unseen conditions. We performed numerical experiments by using the Burgers and Navier-Stokes equations with quadratic nonlinearity. Our results show robustness of the proposed approach in building reduced order models for parameterized systems and confirm the improved trade-off between accuracy and efficiency. 
\end{abstract}


\keywords{Hybrid analysis and modeling \and Supervised machine learning \and Long short-term memory \and Model reduction \and Galerkin projection \and Grassmann manifold.} 
\maketitle

\section{Introduction} \label{sec:Intro}
Physical models are often sought because of their reliability, interpretability, and generalizability being derived from basic principles and physical intuition. However, accurate solution of these models for complex systems usually requires the use of very high spatial and temporal resolutions and/or sophisticated discretization techniques. This limits their applications to offline simulations over a few set of parameters and short time intervals since they can be excessively computationally-demanding. Although those are valuable in understanding physical phenomena and gaining more insight, realistic applications often require near real-time and multi-query responses \cite{singh2018engineering}. Therefore, cheaper numerical approximations using ``adequate-fidelity'' models are usually acceptable \cite{arcucci2019optimal}. In this regard, reduced order modeling offers a viable technique to address systems characterized by underlying patterns \cite{bai2002krylov,lucia2004reduced,hess2019localized,kramer2019nonlinear,swischuk2019projection,bouvrie2017kernel,hamzi2019local,korda2018data,korda2018linear,peitz2019multiobjective}. This is especially true for fluid flows dominated by coherent structures (e.g., atmospheric and oceanic flows) \cite{holmes2012turbulence,taira2017modal,taira2019modal,noack2011reduced,rowley2017model,nair2019transported,kaiser2014cluster,haasdonk2011training,dihlmann2015certified}.

Reduced order models (ROMs) have shown great success for prototypical problems in different fields. In particular, Galerkin projection (GP) coupled with proper orthogonal decomposition (POD) capability to extract the most energetic modes has been used to build ROMs for linear and nonlinear systems \cite{ito1998reduced,iollo2000stability,rowley2004model,milk2016pymor,puzyrev2019pyrom,bergmann2009enablers,couplet2005calibrated,kunisch2001galerkin}. These ROMs preserve sufficient interpretability and generalizability since they are constructed by projecting the full order model (FOM) operators (from governing equations) on a reduced subspace. Despite that, Galerkin projection ROMs (GROMs) have severe limitations in practice, especially for systems with strong nonlinearity. Most fluid flows exhibit quadratic nonlinearity, which makes the computational cost of the resulting GROMs $\sim O(R^3)$, where $R$ is the number of retained modes in ROM approximation. As a result, $R$ should be kept as low as possible (e.g., $O(5)$) through modal truncation for practical purposes; however, this has two main consequences. First, the solution is enforced to live in a smaller subspace which might not contain enough information to accurately represent complex realistic systems. Examples include advection-dominated flows and parametric systems where the decay of Kolmogorov $n$-width is slow \cite{kolmogoroff1936uber,pinkus2012n}. Second, due to the inherent system's nonlinearity, the truncated modes interact with the retained modes. In GROM, these interactions are simply eliminated by modal truncation, which often generates instabilities in the approximation \cite{lassila2014model,rempfer2000low,noack2003hierarchy}. Several efforts have been devoted to introduce stabilization and closure techniques \cite{rahman2019dynamic,sirisup2004spectral,san2014basis,protas2015optimal,cordier2013identification,osth2014need,couplet2003intermodal,kalb2007intrinsic,kalashnikova2010stability,xie2018data,mohebujjaman2019physically,wang2012proper,akhtar2012new,balajewicz2012stabilization,amsallem2012stabilization,san2015stabilized,gunzburger2019evolve} to account for the effects of truncated modes on ROM's dynamics.

In the present study, we aim to address the above problems while preserving considerable interpretability, and generalizability at the core of our uplifted ROM (UROM) approach. In UROM, we present a three-modeling layer framework. In the first layer, we use a standard Galerkin projection method based on the governing equations to model the large scales of the flow (represented by the first few $R$ POD modes) and provide a predictor for the temporal evolution. In the next layer, we introduce a corrector step to correct the Galerkin approximation and make up for the interactions of the truncated modes (or scales) with the large ones. These large scales contribute most to the total system's energy. That is why we dedicate two layers of our approach to correctly resolve them, one of which (i.e., the Galerkin projection) is totally physics-based to promote framework generality. In the third layer, we uplift our approximation and extend the solution subspace to recover some of the flow's finer details by learning a map between the large scales (predicted using the first two layers) and smaller scales.

In particular, we choose the first $R \approx O(5)$ modes to represent the resolved largest scales and the next $Q-R$ modes to represent the resolved smaller scales, where $Q$ is about $4$ to $8$ times larger than $R$. For the second and third layers, we incorporate memory embedding through the use of long short-term memory (LSTM) neural network architecture \cite{hochreiter1997long,Gers99learningto}. Machine learning (ML) tools (of which neural networks is a subclass) have been gaining popularity in fluid mechanics community and analysis of dynamical systems \cite{brunton2019machine,kutz2017deep,durbin2018some,duraisamy2019turbulence,gamboa2017deep,lusch2018deep,otto2019linearly,lee2019model}. In particular, LSTMs have shown great success in learning maps of sequential data and time-series \cite{yeo2019deep,jaeger2004harnessing,lecun2015deep,vlachas2018data,sak2014long}. It should be noted here that UROM can be thought of as a way of augmenting physical models with data-driven tools and vice versa. For the former, an LSTM closure model (second component) is developed to correct GROM and an LSTM super-resolution model (third component) is constructed to uplift GROM and resolve smaller scales. This relaxes the computational cost of GROM to account only for a few $R$ modes. On the other hand, LSTMs in UROM framework are fed with inputs coming from a physics-based approach. This is one way of utilizing physical information rather than fully depend on ML results.

To illustrate the UROM framework, we consider two convection-dominated flows as test cases. The first is the one-dimensional (1D) Burgers equation, which is a simplified benchmark problem for fluid flows with strong nonlinearity. As the second test case, we investigate the two-dimensional (2D) Navier-Stokes equations for a flow with interacting vortices, namely the vortex merger. We compare the UROM approach with the standard GROM approach using $R$ and $Q$ modes. We also investigate a fully non-intrusive ROM (NIROM) approach in these flow problems. We perform a comparison in terms of solution accuracy and computational time to show the pros and cons of UROM with respect to either GROM or NIROM approaches. The rest of the paper is outlined here. In Section~\ref{sec:POD}, we introduce the POD technique for data compression and constructing lower-dimensional subspaces to approximate the solution. For an out-of-sample control parameter (e.g., Reynolds number), we describe basis interpolation via Grassmann manifold approach in Section~\ref{sec:Grassmann}. As a standard physics-informed technique for building ROMs, we introduce Galerkin projection in Section~\ref{sec:GP} with a brief description of the governing equations of the test cases as well as their corresponding GROMs. In Section~\ref{sec:UROM}, we present the proposed UROM framework with a description of its main features. We give results and corresponding discussions in Section~\ref{sec:Results}. Finally, we draw concluding remarks and insights in Section~\ref{sec:Conc}.

\section{Proper Orthogonal Decomposition} \label{sec:POD}
Proper orthogonal decomposition (POD) is one of the most popular techniques for dimensionality reduction and data compression \cite{sirovich1987turbulence,berkooz1993proper,holmes2012turbulence,chatterjee2000introduction,rathinam2003new}. Given datasets, POD provides a linear subspace that minimizes the projection error between the true data and its projection compared to all possible linear subspaces with the same dimension. If a number of $N_s$ data snapshots, $\mathbf{u}(\mathbf{x},t_n)$, where $n \in \{1,2,\dots, N_s \}$,  $\mathbf{x} \in \mathbb{R}^N$ ($N$ being the spatial resolution), are collected in a snapshot matrix $\mathbf{A} \in \mathbb{R}^{N \times N_s}$, then a reduced (or thin) singular value decomposition (SVD) can be applied to this matrix as
\begin{align}
    \mathbf{A} = \mathbf{U} \mathbf{\Sigma} \mathbf{V}^T,
\end{align}
where $\mathbf{U} \in \mathbb{R}^{N \times N_s}$ is a unitary matrix whose columns are the left-singular vectors of $\mathbf{A}$, also known as spatial basis and $\mathbf{V} \in \mathbb{R}^{N_s \times N_s}$ is also a unitary matrix whose columns represent the right-singular vectors, sometimes referred to as temporal basis . $\mathbf{\Sigma} \in \mathbb{R}^{N_s \times N_s}$ is a diagonal matrix whose entries are the singular values of $\mathbf{A}$ (square-roots of the largest $N_s$ eigenvalues of  $\mathbf{A} \mathbf{A}^T$ or $\mathbf{A}^T \mathbf{A}$). In $\mathbf{\Sigma}$, the singular values $\sigma_i$ are sorted in a descending order such that $\sigma_1 \ge \sigma_2 \ge \dots \ge \sigma_{N_s} \ge 0$.

For dimensionality reduction purposes, only the first $R$ columns of $\mathbf{U}$ (denoted as $\widehat{\mathbf{U}}$), the first $R$ columns of $\mathbf{V}$ (denoted as $\widehat{\mathbf{V}}$), and the upper-left $R\times R$ block sub-matrix of $\mathbf{\Sigma}$ (denoted as $\widehat{\mathbf{\Sigma}}$) are retained to provide a reduced order approximation $\widehat{\mathbf{A}}$ of $\mathbf{A}$ as
\begin{align}
    \widehat{\mathbf{A}} = \widehat{\mathbf{U}} \widehat{\mathbf{\Sigma}} \widehat{\mathbf{V}}^T.
\end{align}
It can be easily shown that this approximation $\widehat{\mathbf{A}}$ satisfies the following infimum \cite{trefethen1997numerical}
\begin{align}
       \| \mathbf{A} - \widehat{\mathbf{A}} \|_2 &= \inf_{ \substack{  \mathbf{B} \in \mathbb{R}^{N \times N_s} \\ \\  rank(\mathbf{B}) \le R  } } \| \mathbf{A} - \mathbf{B} \|_2 \label{eq:svd1}\\
        \| \mathbf{A} - \widehat{\mathbf{A}} \|_2 &= \sigma_{R+1}, \label{eq:svd2}
\end{align}
where $\| \cdot \|_2$ refers to the induced $\ell_2$ matrix norm. Equation~\ref{eq:svd1} means that across all possible matrices $\mathbf{B} \in \mathbb{R}^{N\times N_s}$ with a rank of $R$ (or less), $\widehat{\mathbf{A}}$ provides the closest one to $\mathbf{A}$ in the $\ell_2$ sense. Moreover, the singular values $\sigma$ provide a measure of the quality of this approximation as Equation~\ref{eq:svd2} shows that the $\ell_2$ norm between the matrix $\mathbf{A}$ and its $R$-rank approximation equals to $\sigma_{R+1}$. From now on, the first $R$ columns of $\mathbf{U}$ will be referred to as the POD modes or basis function, denoted as $\mathbf{\Phi} = [\phi_1, \phi_2, \dots, \phi_R]$. 


\section{Grassmann Manifold Interpolation} \label{sec:Grassmann}
In recent years, Grassmann manifold has attracted great interest in various applications including model order reduction for parametric systems \cite{amsallem2008interpolation,amsallem2009method,zimmermann2018geometric,zimmermann2019manifold,oulghelou2018fast}. The Grassmann manifold, ${\cal{G}}(q,N)$, is a set of all $q$-dimensional subspaces in $\mathbb{R}^N$, where $0 \le q\le N$. A point $[\mathbf{\Phi}] \in {\cal{G}}(q,N) $ is given as  \cite{oulghelou2018non}
\begin{equation}
    [\mathbf{\Phi}] = \{ \mathbf{\Phi} Q \ \big| \ \mathbf{\Phi}^T \mathbf{\Phi} = I_q, Q\in {\cal{O}}(q)\},
\end{equation}
where $\mathbf{\Phi} \in \mathbb{R}^{N\times q}$ and ${\cal{O}}(q)$ is the group of all $q \times q$ orthogonal matrices. This point represents a $q$-dimensional subspace ${\cal{S}}$ in $\mathbb{R}^N$ spanned by the columns of $\mathbf{\Phi}$. At each point $[\mathbf{\Phi}] \in {\cal{G}}(q,N) $, a tangent space ${\cal{T}}([\mathbf{\Phi}])$ of the same dimension, $N \times q$, can be defined as follows \cite{edelman1998geometry,absil2004riemannian}
\begin{equation}
    {\cal{T}}([\mathbf{\Phi}]) = \{ {\cal{X}} \in \mathbb{R}^{N\times q} \ \big| \ \mathbf{\Phi}^T {\cal{X}} = \mathbf{0} \}. \label{eq:GrassTan1}
\end{equation}
Similarly, each point $[\mathbf{\Gamma}]$ on ${\cal{T}}$ represents a subspace spanned by the columns of $\mathbf{\Gamma}$. This tangent space is a vector space with its origin at $[\mathbf{\Phi}]$. An exponential mapping from a point $[\mathbf{\Gamma}] \in {\cal{T}}([\mathbf{\Phi}])$ to $[\mathbf{\Psi}] \in {\cal{G}}(q,N) $ can be defined as
\begin{align}
    \mathbf{\Psi} = \bigg (\mathbf{\Phi} \mathbf{V} \cos{(\mathbf{\Sigma})} + \mathbf{U} \sin{(\mathbf{\Sigma})} \bigg ) \mathbf{V}^T \label{eq:GrassExp},
\end{align}
where $\mathbf{U}$, $\mathbf{\Sigma}$, $\mathbf{V}$ are obtained from the reduced SVD of $\mathbf{\Gamma}$ as $\mathbf{\Gamma} = \mathbf{U} \mathbf{\Sigma} \mathbf{V}^T$. Inversely, a logarithmic map can be defined from a point $[\mathbf{\Psi}]$ in the neighborhood of $[\mathbf{\Phi}]$ to $[\mathbf{\Gamma}] \in {\cal{T}}([\mathbf{\Phi}])$ is defined as
\begin{align}
    \mathbf{\Gamma} = \mathbf{U} \tan^{-1}{(\mathbf{\Sigma})} \mathbf{V}^T \label{eq:GrassLog},
\end{align}
where $(\mathbf{\Psi} - \mathbf{\Phi} \mathbf{\Phi}^T  \mathbf{\Psi} ) (\mathbf{\Phi} \mathbf{\Psi})^{-1} = \mathbf{U} \mathbf{\Sigma} \mathbf{V}^T$. We would like to note here that the trigonometric functions are applied element-wise on the diagonal entries.

To demonstrate the procedure in our ROM context, for a number $N_p$ of control parameters $\{\mu_i\}_{i=1}^{N_p}$, different sets of POD basis functions are computed corresponding to each parameter, denoted as $\{ \mathbf{\Phi}_{i} \}_{i=1}^{N_p}$. These bases correspond to a set of points on the Grassmann manifold ${\cal{G}}(R,N) $. To perform an out-of-sample testing, the basis functions $\mathbf{\Phi}_{\text{Test}}$ for the test parameter $\mu_{\text{Test}}$ should be computed through interpolation. However, direct interpolation of the POD bases is not effective since it is an interpolation on a non-flat space and it does not guarantee that the resulting point would lie on ${\cal{G}}(R,N)$. Moreover, the optimality and orthonormality characteristics of POD are not necessarily conserved. Alternatively, the tangent space ${\cal{T}}$ is a flat space where standard interpolation can be performed effectively. First, a reference point at the Grassmann manifold is selected, corresponding to $\mathbf{\Phi}_{\text{Ref}}$. The tangent plane at this point is thus defined using Equation~\ref{eq:GrassTan1}. Then, the neighboring points on Grassmann manifold corresponding to the subspaces spanned by $\{ \mathbf{\Phi}_{i} \}_{i=1}^{N_p}$ are mapped onto that tangent plane using the logarithmic map, defined in Equation ~\ref{eq:GrassLog} to calculate $\{ \mathbf{\Gamma}_{i} \}_{i=1}^{N_p}$. Standard Lagrange interpolation can be performed to compute $ \mathbf{\Gamma}_{\text{Test}}$ as follows
\begin{align}
    \mathbf{\Gamma}_{\text{Test}} =  \mathlarger{\sum}_{i=1}^{N_p} { \bigg( \mathlarger{\prod}_{\substack{ j=1 \\ j \ne i}}^{N_p} \dfrac{\mu_{\text{Test}}-\mu_{j}}{\mu_{i}-\mu_{j}} \bigg) \mathbf{\Gamma}_{i} }.
\end{align}
Finally, the point $[\mathbf{\Gamma}_{\text{Test}}] \in {\cal{T}}([\mathbf{\Phi}_{\text{Ref}}])$ is mapped to the Grassmann manifold ${\cal{G}}(R,N)$ to obtain the set of POD basis functions at the test parameter, $\mathbf{\Phi}_{\text{Test}}$, using the exponential map given in Equation~\ref{eq:GrassExp}. Hence, an interpolation on the tangent plane to Grassmann manifold provides a basis of the same dimension (i.e., $[\mathbf{\Phi}_{\text{Test}}] \in {\cal{G}}(R,N)$). Moreover, it preserves the orthonormality of the basis (i.e., the columns of $\mathbf{\Phi}_{\text{Test}}$ are orthonormal to each other). Those properties are not guaranteed if standard interpolation techniques are used directly to inpterpolate basis. The procedure for Grassmann manifold interpolation is summarized in Algorithm~\ref{alg:Grass}.

\begin{breakablealgorithm}
  \caption{Grassmann manifold interpolation}
  \label{alg:Grass}
  \begin{algorithmic}[1]
    \State Given a set of basis functions $\mathbf{\Phi}_1, \mathbf{\Phi}_2, \dots, \mathbf{\Phi}_P$ corresponding to the offline simulations parameterized by $\mu_1, \mu_2, \dots, \mu_P$.
    \State Select a point $[\Phi_{\text{Ref}}] \leftarrow [\Phi_{i}] \in \{ [\Phi_{1}], \dots, [\Phi_{N_p}] \}$ corresponding to the basis function set $\mathbf{\Phi}_{\text{Ref}} \leftarrow \mathbf{\Phi}_i \in [\mathbf{\Phi}_1, \dots, \mathbf{\Phi}_{N_p}]$ as the reference point.
    \State Map each point $[\Phi_{i}] \in {\cal{G}}(N,R)$ to $[\mathbf{\Gamma}_i] \in {\cal{T}}([\mathbf{\Phi}_{\text{Ref}}])$ using logarithmic map 
    \begin{equation}
    (\mathbf{\Phi}_i-\mathbf{\Phi}_{\text{Ref}} \mathbf{\Phi}_{\text{Ref}}^T \mathbf{\Phi}_i)(\mathbf{\Phi}_{\text{Ref}}^T\mathbf{\Phi}_i)^{-1} = \mathbf{U}_i\mathbf{\Sigma}_i \mathbf{V}_i^T,
    \end{equation}
    \begin{equation}
    \mathbf{\Gamma}_i =  \mathbf{U}_i \text{tan}^{-1}(\mathbf{\Sigma}_i) \mathbf{V}_i^T.
    \end{equation}
    \State Construct matrix $\mathbf{\Gamma}_{\text{Test}}$ corresponding to the test parameter $\mu_{\text{Test}}$ using Lagrange interpolation of matrices $\mathbf{\Gamma}_i$, corresponding to $\mu_1,\dots,\mu_{N_p}$
    \begin{equation}
    \mathbf{\Gamma}_{\text{Test}} = \sum_{i=1}^{N_p}\bigg( \prod_{\substack{j=1 \\ j\neq i}}^{P}\frac{\mu_{\text{Test}} - \mu_j}{\mu_i - \mu_j}\bigg)\mathbf{\Gamma}_i.
    \end{equation}
    \State Compute the POD basis functions $\mathbf{\Phi_{\text{Test}}}$ corresponding to the test parameter $\mu_{\text{Test}}$ using the exponential map
    \begin{equation}
    \mathbf{\Gamma}_{\text{Test}} = \mathbf{U}_{\text{Test}} \mathbf{\Sigma}_{\text{Test}} \mathbf{V}_{\text{Test}}^{T},
    \end{equation}
    \begin{equation}
    \mathbf{\Phi}_{\text{Test}} = \bigg ( \mathbf{\Phi}_{\text{Ref}} \mathbf{V}_{\text{Test}} \cos(\mathbf{\Sigma}_{\text{Test}}) + \mathbf{U}_{\text{Test}} \sin(\mathbf{\Sigma}_{\text{Test}}) \bigg )  \mathbf{V}_{\text{Test}}^{T},
    \end{equation}
  where the trigonometric operators apply only to the diagonal elements.
  \end{algorithmic}
\end{breakablealgorithm}

\section{Galerkin Projection} \label{sec:GP}

To emulate the system's dynamics in ROM context, a Galerkin projection is usually performed. In Galerkin projection-based ROM (GROM), the solution $\mathbf{u}(\mathbf{x},t_n)$ is approximated by $\mathbf{\hat{u}}(\mathbf{x},t_n)$. First, $\mathbf{\hat{u}}(\mathbf{x},t_n)$ is constrained to lie in a trial subspace ${\cal{S}}$ spanned by the basis $\mathbf{\Phi}$. In our study, this basis is computed using the POD method presented in Section~\ref{sec:POD}. Then, the full-order operators are projected onto the same subspace ${\cal{S}}$. In other words, the residual of the governing ODE is enforced to be orthogonal to ${\cal{S}}$. Galerkin projection can be viewed as a special case of Petrov-Galerkin method \cite{carlberg2011efficient,carlberg2017galerkin,choi2019space,xiao2013non} considering the same trial subspace as a test subspace. In the following, we present the governing equations for our test cases, namely 1D Burgers equation and 2D Navier-Stokes equations as well as their low-order approximations.

\subsection{1D Burgers Equation} \label{sec:burg}
The one-dimensional (1D) viscous Burgers equation represents a standard benchmark for the analysis of nonlinear advection-diffusion problems in 1D setting with similar quadratic nonlinear interaction and Laplacian dissipation. The evolution of the velocity field $u(x, t)$, in a dimensionless form, is given by
\begin{equation}
    \dfrac{\partial u}{\partial t} + u \dfrac{\partial u}{\partial x} = \dfrac{1}{\text{Re}} \dfrac{\partial ^2 u}{\partial x^2}, \label{eq:brg}
\end{equation}
where $\text{Re}$ is the dimensionless Reynolds number, defined as the ratio of inertial effects to viscous effects. Equation~\ref{eq:brg} can be rewritten as
\begin{align} \label{eq:burg_semi}
    \dfrac{\partial {u}}{\partial t} &= \dfrac{1}{\text{Re}} \dfrac{\partial^2 u}{\partial x^2}  - u \dfrac{\partial u}{\partial x}.
\end{align}
Then, the reduced-rank approximation  $\hat{u}(x,t) = \sum_{k=1}^R \alpha_k(t) \phi_k(x)$ (where $\phi_k$ are the constructed POD modes and  $\alpha_k$ are the corresponding coefficients) is plugged into this equation and an inner product with an arbitrary basis $\phi_k$ is performed to give the following dynamical ODE which is the GROM for the Burgers equation
\begin{align}
    \dfrac{\text{d}\alpha_k}{\text{d}t} &=  \sum_{i=1}^{R} \mathfrak{L}_{i,k} \alpha_i + \sum_{i=1}^{R} \sum_{j=1}^{R} \mathfrak{N}_{i,j,k} \alpha_i \alpha_j,  \qquad k=1,2,\dots,R,\label{eq:rombrg}
\end{align}
where $\mathfrak{L}$ and $\mathfrak{N}$ are the matrix and tensor of predetermined model coefficients corresponding to linear and nonlinear terms, respectively. They are precomputed as
\begin{align*}
    \mathfrak{L}_{i,k} & = \big\langle \dfrac{1}{\text{Re}} \dfrac{\partial^2 \phi_i }{\partial x^2}  ;  \phi_k \big\rangle, \\
    \mathfrak{N}_{i,j,k} &= \big\langle -\phi_i \dfrac{\partial \phi_j}{\partial x};  \phi_k \big\rangle,
\end{align*}
where the angle-parentheses refer to the Euclidean inner product defined as $\langle \mathbf{x} , \mathbf{y} \rangle = \mathbf{x}^T \mathbf{y} =  \sum_{i=1}^{N} x_iy_i$. 
\subsection{2D Navier-Stokes Equations} \label{sec:NS}
The vorticity-streamfunction formulation of the two-dimensional (2D) Navier-Stokes equations can be written as \cite{Gun89}
\begin{align} 
\dfrac{\partial \omega}{\partial t} + J(\omega,\psi) &= \dfrac{1}{\text{Re}} \nabla^2 \omega, \label{eq:NS}
\end{align}
where $\omega$ is the vorticity and $\psi$ is the streamfunction. The vorticity-streamfunction formulation prevents the odd-even decoupling issues that might arise between pressure and velocity components and enforces the incompressibility condition. The kinematic relationship between vorticity and streamfunction is given by the following Poisson equation,
\begin{equation}\label{eq:Poisson}
\nabla^2 \psi = -\omega.
\end{equation}
Equation~\ref{eq:NS} and Equation~\ref{eq:Poisson} include two operators, the Jacobian ($J(f,g)$) and the Laplacian ($\nabla^2 f$) defined as
\begin{align}
    J(f,g) &= \dfrac{\partial f}{\partial x} \dfrac{\partial g}{\partial y} -  \dfrac{\partial f}{\partial y} \dfrac{\partial g}{\partial x}, \\
    \nabla^2 f &= \dfrac{\partial^2 f}{\partial x^2} + \dfrac{\partial^2 f}{\partial y^2}.
\end{align}
Similar to the 1D Burgers problem, Equation~\ref{eq:NS} can be rearranged as
\begin{align} \label{eq:NS_semi}
    \dfrac{\partial {\omega}}{\partial t} &= \dfrac{1}{\text{Re}} \nabla^2 \omega - J(\omega,\psi).
\end{align}
The reduced-rank approximations of the vorticity and streamfunction fields can be written as follows
\begin{align}
    \omega(x,y,t) &=  \sum_{k=1}^R \alpha_k(t) \phi_k^{\omega}(x,y), \\
    \psi(x,y,t_n) &=  \sum_{k=1}^R \alpha_k(t) \phi_k^{\psi}(x,y).
\end{align}
We note that the vorticity and streamfunction share the same time-dependent coefficients ($\alpha_k(t)$) since they are related through the kinematic relationship given by Equation~\ref{eq:Poisson} (i.e., streamfunction is not a prognostic variable). Moreover, as POD preserves linear properties, the spatial POD modes for streamfunction can be obtained from the vorticity modes by solving the following Poisson equations
\begin{align}
    \nabla^2 \phi_k^{\psi}(x,y) = -\phi_k^{\omega}(x,y), \quad k=1,2,\dots, R.
\end{align}
The GROM for the 2D Navier-Stokes equations is given by the same ODE (Equation~\ref{eq:rombrg}) with the following coefficients
\begin{align*}
    \mathfrak{L}_{i,k} &= \big\langle  \dfrac{1}{\text{Re}} \nabla^2 \phi_i^{\omega} ; \phi_k^{\omega} \big\rangle, \\
    \mathfrak{N}_{i,j,k} &= \big\langle -J(\phi_i^{\omega},\phi_j^{\psi}) ; \phi_k^{\omega} \big\rangle.
\end{align*}

\begin{figure}[!ht]
	\centering
	\includegraphics[trim= 0 0 330 0, clip, width=0.95\textwidth]{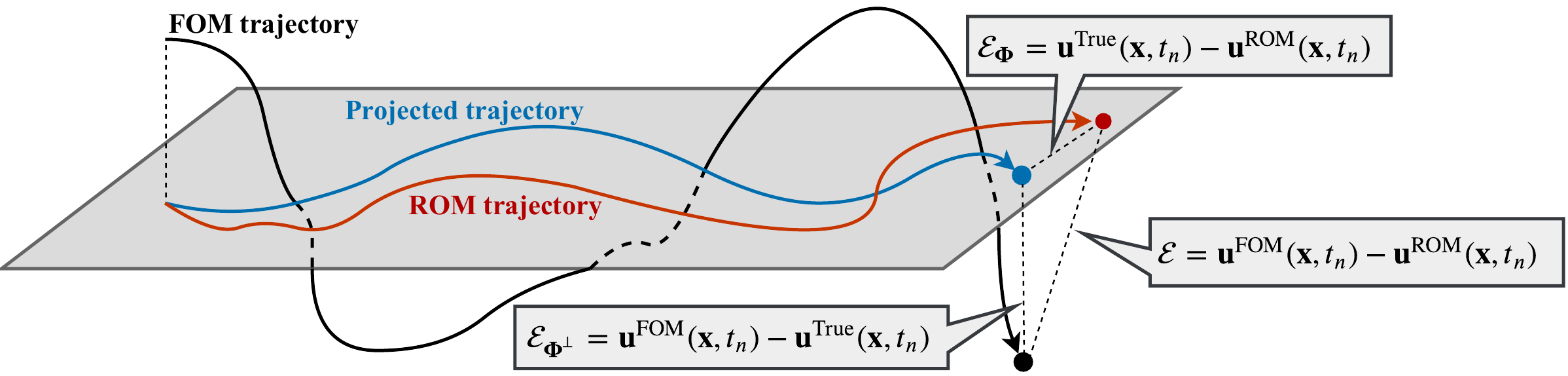}
	\caption{A representation of error sources in ROM (e.g., see  \cite{rathinam2003new,amsallem2010interpolation} for further details).}
	\label{fig:ROMerror}
\end{figure}

Due to the modal truncation and inherent nonlinearity in Equation~\ref{eq:rombrg}, GROM no longer represents the same system (i.e., it solves a different problem). As a result, the obtained trajectory from solving the ROM deviates from the projected trajectory, as shown in Figure~\ref{fig:ROMerror}. Therefore, the optimality of the POD basis is lost. Moreover, due to the triadic nonlinear interactions, instabilities can occur in GROMs. To mitigate these problems, closure and/or stabilization techniques are usually required to obtain accurate results. Increasing the ROM dimension can improve the results. However, due to the nonlinearity of the resulting ROM, the computational cost of GROM is $O(R^3)$, which severely constrains the ROM dimension used in practical applications.

We note here that we are adopting the tensorial GROM approach \cite{stefuanescu2014comparison}, where the coefficients $\mathfrak{L}$ and $\mathfrak{N}$ are computed offline. Other approaches can include online computations while incorporating the empirical interpolation method (EIM) \cite{barrault2004eim} or its discrete version, the discrete empirical interpolation method (DEIM) \cite{chaturantabut2010nonlinear} to reduce the online computational cost.

\section{Uplifted Reduced Order Modeling} \label{sec:UROM}
\begin{figure}[!ht]
	\centering
	\includegraphics[trim= 0 0 240 0, clip, width=0.95\textwidth]{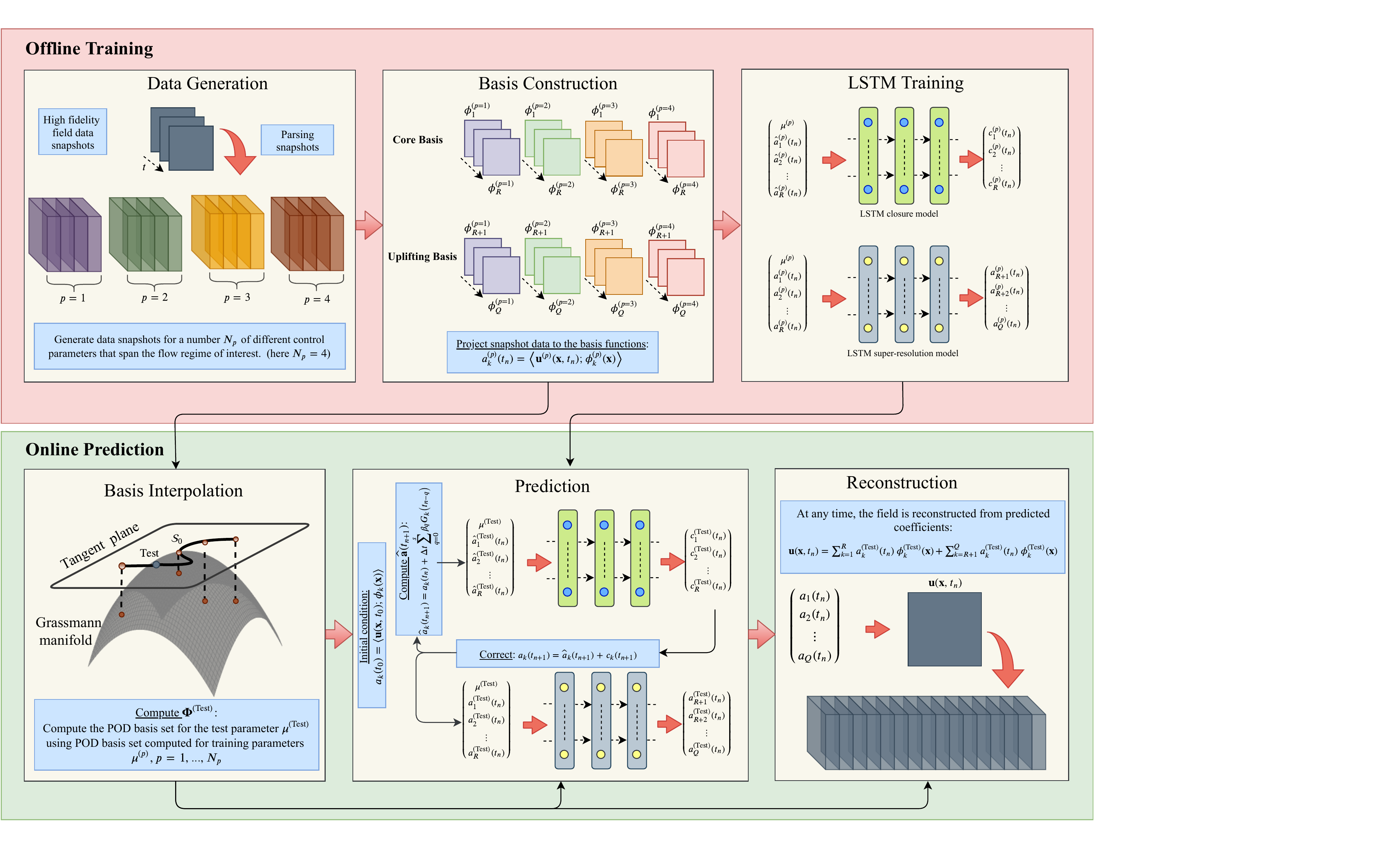}
	\caption{A schematic diagram for the workflow of UROM framework.}
	\label{fig:UROM}
\end{figure}

As noted in Section~\ref{sec:GP}, the computational cost of GROM is $O(R^3)$, which limits the number of modes to be used in the ROM. This modal truncation has two major consequences. First, the flow field variable is constrained to lie in a small subspace, spanned by the very first few modes. For convection-dominated flows or parametric problems characterized by slow decay of the Kolmogorov $n$-width, these few modes may be less representative of the true physical system. This significantly reduces the accuracy of the resulting ROM. This is shown as the projection error ${\cal{E}}_{\mathbf{\Phi}^{\perp}}$ in Figure~\ref{fig:ROMerror}, since the truncated modes are orthogonal to the subspace spanned by $\mathbf{\Phi}$. Second, due to the inherent nonlinearity, the truncated modes (or scales) interact with the retained ones. Thus, this truncation simply ignores these interactions, often giving rise to numerical instabilities of solution. This error is represented as ${\cal{E}}_{\mathbf{\Phi}}$ in Figure~\ref{fig:ROMerror} since it lies in the same subspace $\mathbf{\Phi}$. In our uplifted reduced order modeling (UROM) framework (presented in Figure~\ref{fig:UROM}), we try to address these two problems. 

We extend our reduced-order approximation $\mathbf{\hat{u}}$ to include the first $Q$ modes, where $Q>R$, assuming that the first $R$ modes account for the resolved large scales, and the next $(Q-R)$ modes represent the resolved small scales (while the remaining truncated modes account for the unresolved scales). So, the UROM approximation $\mathbf{\hat{u}}(\mathbf{x},t_n)$ can be expanded as
\begin{align}
    \mathbf{\hat{u}}(\mathbf{x},t_n) = \underbrace{\sum_{k=1}^R a_k(t) \phi_k(\mathbf{x})}_{\substack{\text{core}\\ \text{(resolved large scales)}}} + \underbrace{\sum_{k=R+1}^Q a_k(t) \phi_k(\mathbf{x})}_{\substack{\text{uplift}\\ \text{(resolved small scales)}}},
\end{align}
where $\mathbf{u}$ is a general notation for the flow field of interest, $\phi_k$ denotes the POD modes, and $a_k$ are the corresponding temporal coefficients. Then, standard Galerkin projection is applied to resolve the large scales (first $R$ modes) to obtain a predictor $\{\hat{a}_k\}_{k=1}^{R}$ as
\begin{align}
    \dfrac{\text{d}\hat{a}_k}{\text{d}t} &= G_k,  \\
    \hat{a}_k(t_{n+1}) &= {a}_k(t_{n}) + {\Delta t} \sum_{q=0}^{s} \beta_q G_k(t_{n-q}),
\end{align}
where $s$ and $\beta_q$ depend upon the numerical scheme used for the time integration. In the present study, we use the third-order Adams-Bashforth (AB3) method for which $s=2,~ \beta_0=23/12,~ \beta_1=-16/12,$ and $\beta_2=5/12$. $G_k$ is obtained by Galerkin projection (discussed in Section~\ref{sec:GP}) as
\begin{align}
    G_k(t_{n}) = \sum_{i=1}^{R} \mathfrak{L}_{i,k} a_i(t_{n}) + \sum_{i=1}^{R} \sum_{j=1}^{R} \mathfrak{N}_{i,j,k} a_i(t_{n}) a_j(t_{n}).
\end{align}

In order to correct GROM results for the first $R$ modes, closure and/or stabilization are required. In our framework, we propose the use of LSTM architecture to learn a correction term to steer the GROM prediction of the modal coefficients $\{\hat{a}_k(t_n)\}_{k=1}^{R}$ to the true values $\{a_k(t_n)\}_{k=1}^{R}$ at each timestep. In other words, an LSTM is trained to learn the map from $\{\hat{a}_k(t_n)\}_{k=1}^{R}$ as input to $\{c_k(t_n)\}_{k=1}^{R}$ as output, where $c$ is a correction (closure) term  defined as
\begin{align}
c_k(t_n)=a_k(t_n)-\hat{a}_k(t_n).
\end{align}
It should be noted here that the introduced data driven closure takes into account the interactions of \emph{all} the fine scales ($k=R+1,\dots,N_s$) with the resolved large scales ($k=1,\dots,R$), as manifested in the data snapshots. Finally, to account for small scales, we train a second super-resolution LSTM neural network to predict the modal coefficients of the next ($Q-R$) modes, where the input of the LSTM is $\{a_k(t_n)\}_{k=1}^{R}$ and the output is $\{a_k(t_{n})\}_{k=R+1}^{Q}$. To improve parametric performance of the UROM architecture and promote generality, the LSTMs' inputs are augmented with the control parameter. Therefore, the LSTM maps $f$ and $g$ corresponding to the closure and super-resolution models, respectively, can be written as
\begin{align}
    f: \begin{bmatrix} \mu \\ \hat{a}_1(t_n) \\ \vdots \\ \hat{a}_R(t_n)  \end{bmatrix} \mapsto \begin{bmatrix} c_1(t_n) \\ \vdots \\ c_R(t_n)      \end{bmatrix} , \qquad
    g: \begin{bmatrix} \mu \\ a_1(t_n) \\ \vdots \\ a_R(t_n)  \end{bmatrix} \mapsto 
    \begin{bmatrix} a_{R+1}(t_n) \\ \vdots \\ a_Q(t_n)      \end{bmatrix} .
\end{align}
In brief, we first steer the red line in Figure~\ref{fig:ROMerror} to the blue one (i.e., introduce data-driven closure by LSTM). Then, we reduce the projection error (difference between the the blue and black lines) by expanding our solution subspace to span $Q$ modes rather than only $R$. Figure~\ref{fig:UROM} demonstrates the building blocks and workflow of our proposed UROM framework in both the offline and online phases. Few merits of the proposed UROM approach can be listed as follows.
\begin{itemize}
    \item The physics-constrained GROM is maintained to account for the large scales. This enriches the framework interpretability and generalizability across a wide range of control parameters.
    \item GROM acts on a few modes, minimizing the online computational cost (i.e., $O(R^3)$, where $R<Q$).
    \item GROM, being physics-informed, can be used as a sanity check to decide whether or not the LSTM predictions should be considered.
    \item Data-driven closure/correction encapsulates information from \emph{all} interacting modes and mechanisms.
    \item Since both LSTMs are fed with input from a physics-based approach, UROM can be considered as a way of enforcing physical knowledge to enhance data-driven tools.
    \item LSTMs' inputs are augmented with the control parameter to provide a more accurate mapping (sometimes also called physics-informed mapping).
\end{itemize}

\subsection{Long Short-Term Memory Embedding} \label{sec:LSTM}
To learn the maps $f$ and $g$ in UROM, we incorporate memory embedding through the use of LSTM architecture. LSTM is a variant of recurrent neural networks capable of learning and predicting the temporal dependencies between the given data sequence based on the input information and previously acquired information. Recurrent neural networks have been used successfully in ROM community to enhance standard projection ROMs \cite{kani2019reduced} and build fully non-intrusive ROMs \cite{mohan2018deep,wang2018model,rahman2019non,ahmed2019memory,wiewel2019latent,xiao2019machine}. In the present study, we use LSTMs to augment the standard physics-informed ROM by introducing closure as well as super-resolution data-driven models. We utilize Keras API \cite{chollet2015keras} to build the LSTMs used in our UROM approach. Details about the LSTM architecture can be found in \cite{ahmed2019memory,rahman2019non}. A summary of the adopted hyperparameters is presented in Table~\ref{table:hyper}. We also found that the constructed neural networks are not very sensitive to hyperparameters. Meanwhile, for optimal hyperparameter selection, different techniques (e.g., gridsearch) can be used to tune them.

\begin{table}[htbp]
    \centering
    \caption{A list of hyperparameters utilized to train the LSTM network for all numerical experiments.}
    \begin{tabular}{p{0.44\textwidth}p{0.21\textwidth}p{0.21\textwidth}}
    \hline\noalign{\smallskip}
    Variables & \quad \quad \ 1D Burgers & 2D Navier-Stokes \\
    \noalign{\smallskip}\hline\noalign{\smallskip}
    Number of hidden layers  & \quad \quad \ \ 3 & \ 3 \\
    Number of neurons in each hidden layer  & \quad \quad \ \ 60 & \ 80 \\
    Number of lookbacks & \quad \quad \ \ 3 & \ 3 \\
    Batch size & \quad \quad \ \ 64 & \ 64 \\
    Epochs & \quad \quad \ \ 200 & \ 200 \\
    Activation functions in the LSTM layers & \quad \quad \ \ tanh & \ tanh \\
    Validation data set & \quad \quad \ \ 20$\%$ & \  20$\%$ \\
    Loss function &  \quad \quad \ \ MSE & \ MSE \\
    Optimizer & \quad \quad \ \ ADAM & \ ADAM \\
    Learning rate & \quad \quad \ \ 0.001 & \ 0.001 \\
    First moment decay rate & \quad \quad \ \ 0.9 & \ 0.9 \\
    Second moment decay rate & \quad \quad \ \ 0.999 & \ 0.999 \\
    \noalign{\smallskip}\hline
    \end{tabular}
    \label{table:hyper}
\end{table}

\section{Results} \label{sec:Results}
In order to demonstrate the features and merits of UROM, we present results for the two test cases at out-of-sample control parameters (interpolatory and extrapolatory). For the number of modes, we use $R=4$ for the core dynamics and $Q=16$ for super-resolution. We compare the accuracy of UROM prediction with the FOM results as well as the true projection of the FOM snapshots on the POD subspace (denoted as `True' in our results), where 
\begin{align}
    a_k^{\text{True}}(t_n) &= \big\langle \mathbf{u}(\mathbf{x},t_n) ; \phi_k(\mathbf{x}) \big\rangle, \\
    \mathbf{u}^{\text{True}}(\mathbf{x},t_n) &= \sum_{k=1}^{Q}  a_k^{\text{True}}(t_n) \phi_k(\mathbf{x}).
\end{align}
Since UROM can be considered as a hybrid approach between fully intrusive and fully non-intrusive ROMs, we compare it with standard Galerkin projection ROM with $4$ and $16$ modes, denoted as GROM($4$) and GROM($16$), respectively. Moreover, we show the results of a fully non-intrusive ROM approach using 16 modes (denoted as NIROM). For this NIROM, we use the same LSTM architecture presented in Section~\ref{sec:LSTM} as a time-stepping integrator. In particular, we learn a map between the values of modal coefficients at current timestep and their values at the following timestep. Also, we augment our input with the control parameter to enhance the mapping accuracy. In other words, the NIROM map $h$ can be represented as follows
\begin{align}
    h: \begin{bmatrix} \mu \\ a_1(t_n) \\ \vdots \\ a_Q(t_n)  \end{bmatrix} \mapsto 
    \begin{bmatrix} a_{1}(t_{n+1}) \\ \vdots \\ a_Q(t_{n+1})      \end{bmatrix}.
\end{align}
Finally, we present the CPU time for UROM, GROM(4), GROM(16), and NIROM to demonstrate the computational gain. 

\subsection{1D Burgers Problem}
For 1D Burgers simulation, we consider the initial condition \cite{maleewong2011line}
\begin{equation}
    u(x, 0) = \dfrac{x}{1 + \exp{\left( \dfrac{\text{Re}}{16} \left(  4x^2 - 1 \right)  \right)} },
\end{equation}
with $ x\in [0,1]$. Also, we assume Dirichlet boundary conditions: $u(0, t) = u(1,t) = 0$. The 1D Burgers equation with the above initial and boundary conditions has the following analytic solution representing a traveling wave \cite{maleewong2011line} 
\begin{equation}
    u(x,t) = \frac{\frac{x}{t+1}}{1 + \sqrt{\frac{t+1}{t_0}} \exp \big(\text{Re} \frac{x^2}{4t + 4} \big)},
\end{equation}
where $t_0=\exp(\text{Re}/8)$. For offline training, we obtain solution for different Reynolds numbers ($\text{Re} \in \{200,400,600,800\}$). For each case, we collect $1000$ snapshots for $t \in [0,1]$ (i.e., $\Delta t = 0.001$). For online deployment, we obtain the POD basis at $\text{Re} = 500$ and $\text{Re} = 1000$ using Grassmann manifold interpolation as discussed in Section~\ref{sec:Grassmann}.

\subsubsection{\textbf{Re = 500}: demonstrating interpolatory capability}
\begin{figure}[!ht]
	\centering
	\includegraphics[width=0.95\textwidth]{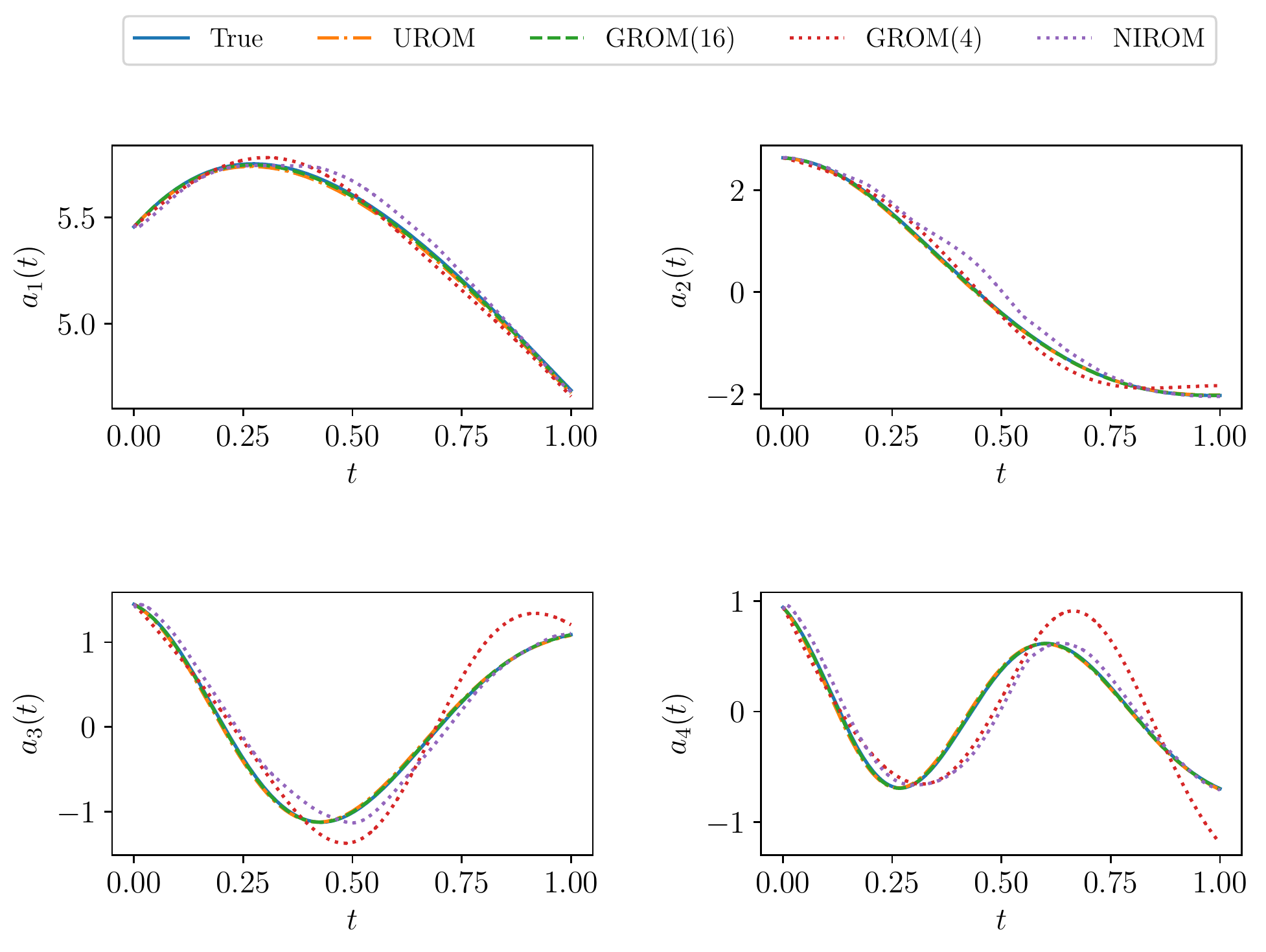}
	\caption{Temporal evolution of the first 4 POD modal coefficients for Burgers problem as predicted by UROM, GROM(4), GROM(16), and NIROM compared with the true values obtained by projection of FOM field on the interpolated modes at $\text{Re} =500$. Note that GROM($4$) and NIROM yield inaccurate results.}
	\label{fig:burg_a500}
\end{figure}

A Reynolds number of $500$ represents an interpolatory case, where we use the POD basis at $\text{Re} = 600$ as our reference point for basis interpolation. The evolution of the first 4 POD temporal coefficients using different frameworks are shown in Figure~\ref{fig:burg_a500}. It is clear that GROM($4$) is incapable of capturing the true dynamics due to the severe modal truncation. On the other hand, both GROM($16$) and UROM show very good results; however, GROM($16$) is more computationally expensive as will be shown in Section~\ref{sec:cpu}.

\begin{figure}[!ht]
	\centering
	\includegraphics[width=0.95\textwidth]{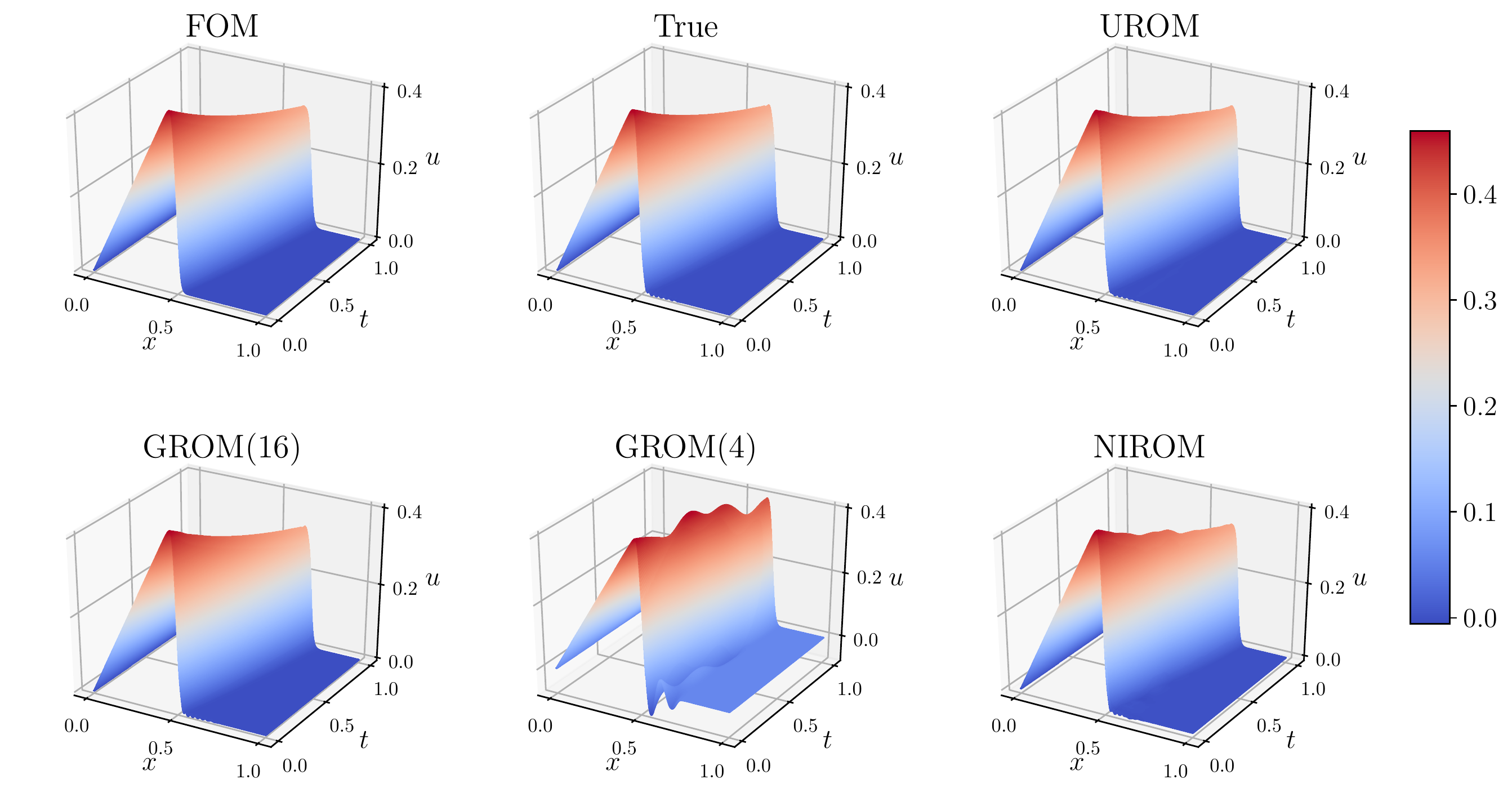}
	\caption{Temporal evolution of velocity fields for Burgers problem at $\text{Re} =500$ with $R=4$ and $Q=16$.}
	\label{fig:u500}
\end{figure}

\begin{figure}[!ht]
	\centering
	\includegraphics[width=0.95\textwidth]{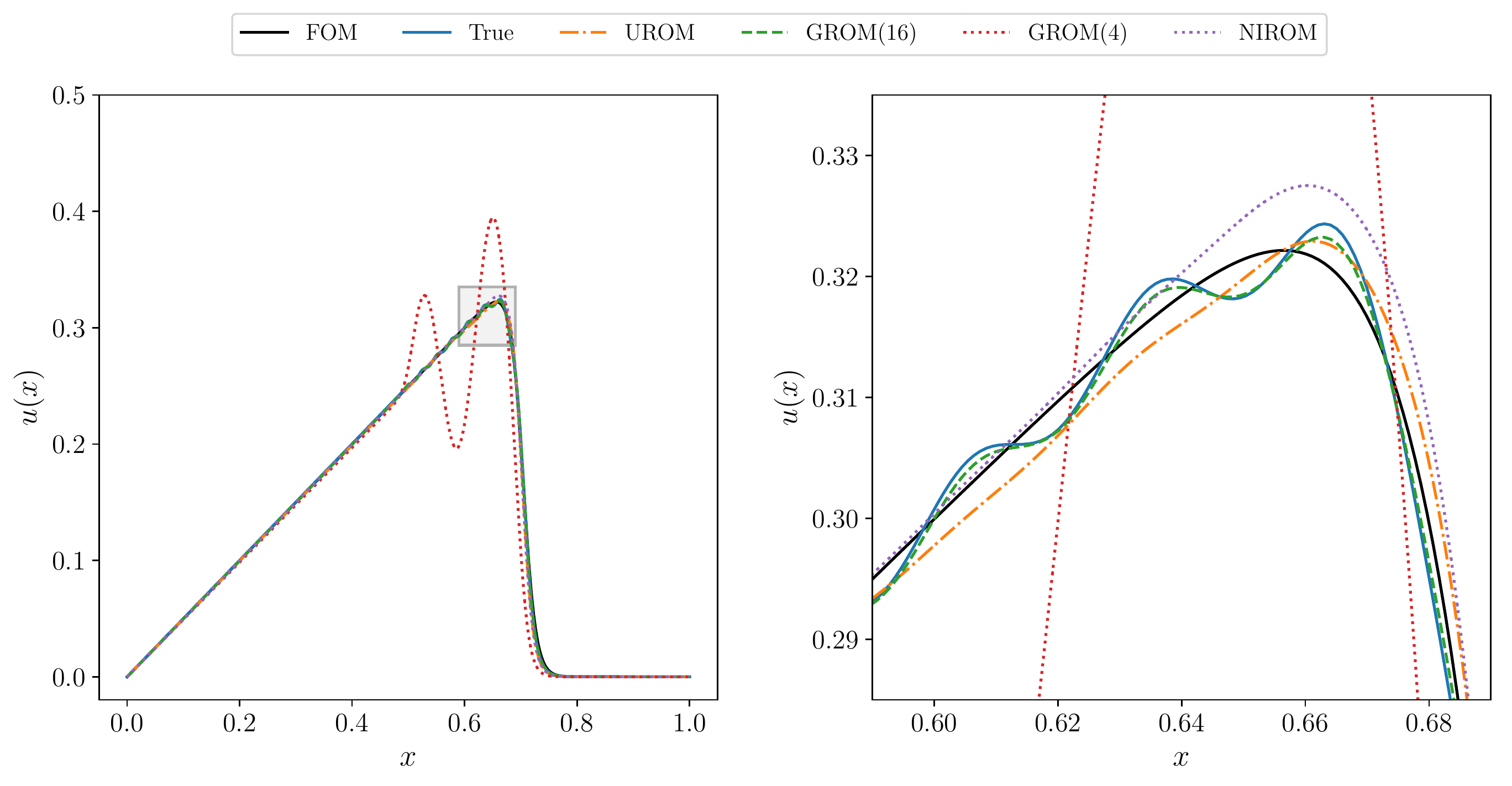}
	\caption{Final velocity fields (at $t=1$) for Burgers problem at $\text{Re} =500$ with a zoom-in view at the right using $R=4$ and $Q=16$. Note that UROM is giving smooth predictions while GROM($4)$ is showing significant oscillations.}
	\label{fig:ufinal500}
\end{figure}

For field reconstruction, we present the temporal field evolution in Figure~\ref{fig:u500} for FOM snapshots, true projection, UROM, GROM, and NIROM. It can be seen that UROM gives very good predictions for field reconstruction compared with GROM($4$) and NIROM, which yield less accurate results. For better visualizations, we show the final field (i.e., at $t=1$) in Figure~\ref{fig:ufinal500} with a close-up view on the region characterizing the wave-shock.

\subsubsection{\textbf{Re = 1000}: demonstrating extrapolatory capability}
\begin{figure}[!ht]
	\centering
	\includegraphics[width=0.95\textwidth]{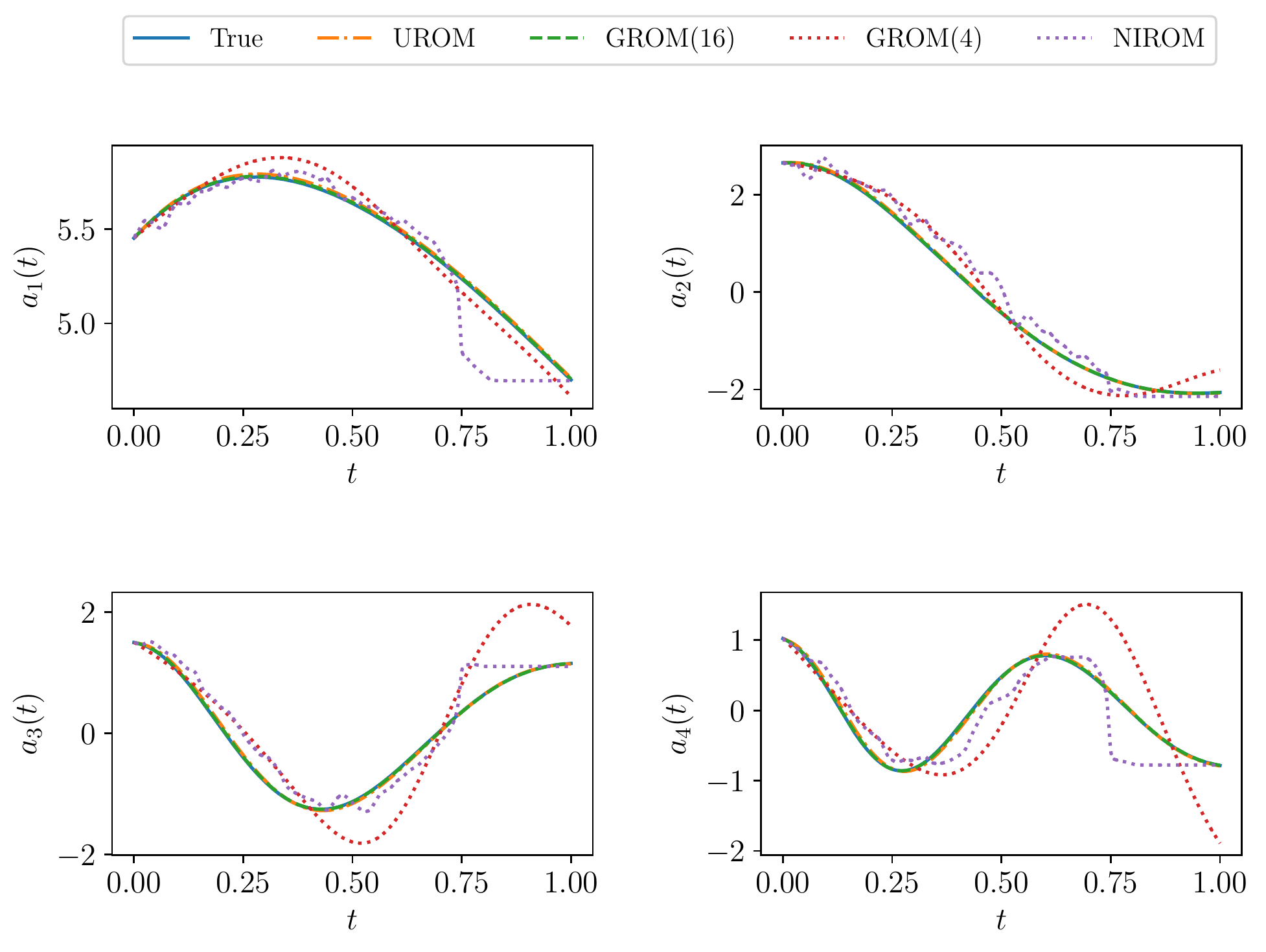}
	\caption{Temporal evolution of the first 4 POD modal coefficients for Burgers problem as predicted by UROM, GROM(4), GROM(16), and NIROM compared with the true values obtained by projection of FOM field on the interpolated modes at $\text{Re} =1000$. GROM($4$) deviates from true trajectory while NIROM gives non-physical predictions.}
	\label{fig:burg_a1000}
\end{figure}

In order to investigate the exptrapolatory performance of UROM, we test the approach at $\text{Re} = 1000$, with the basis at $\text{Re} = 800$ as reference point for interpolation. The POD modal coefficients are shown in Figure~\ref{fig:burg_a1000}, where we can see that both UROM and GROM($16$) are still capable to capture the true projected trajectory. Interestingly, the NIROM predictions are less satisfactory, giving non-physical behavior at some time instants. This suggests that the physical core of UROM promotes its generality, compared to the totally data-driven NIROM approach. However, we note that the deficient behavior of NIROM can be partly due to the sub-optimal architecture for our network as we only use the same hyperparameters (except for the size of input and output layers) for all simulations (as given in Table~\ref{table:hyper}). More sophisticated architectures and further tuning of hyperparameters would probably improve NIROM predictions.

\begin{figure}[!ht]
	\centering
	\includegraphics[width=0.95\textwidth]{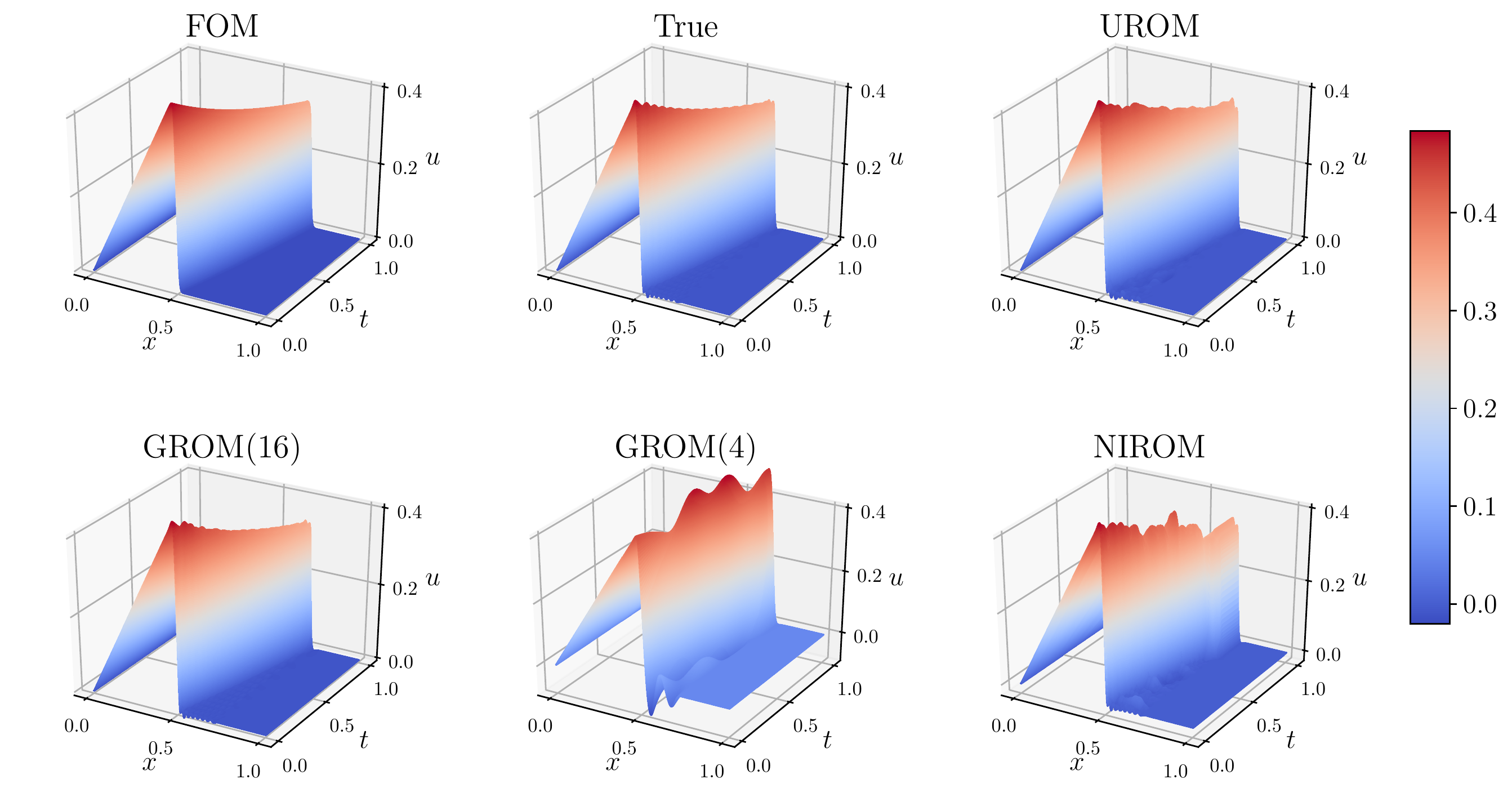}
	\caption{Temporal evolution of velocity fields for Burgers problem at $\text{Re} =1000$ with $R=4$ and $Q=16$.}
	\label{fig:u1000}
\end{figure}

\begin{figure}[!ht]
	\centering
	\includegraphics[width=0.95\textwidth]{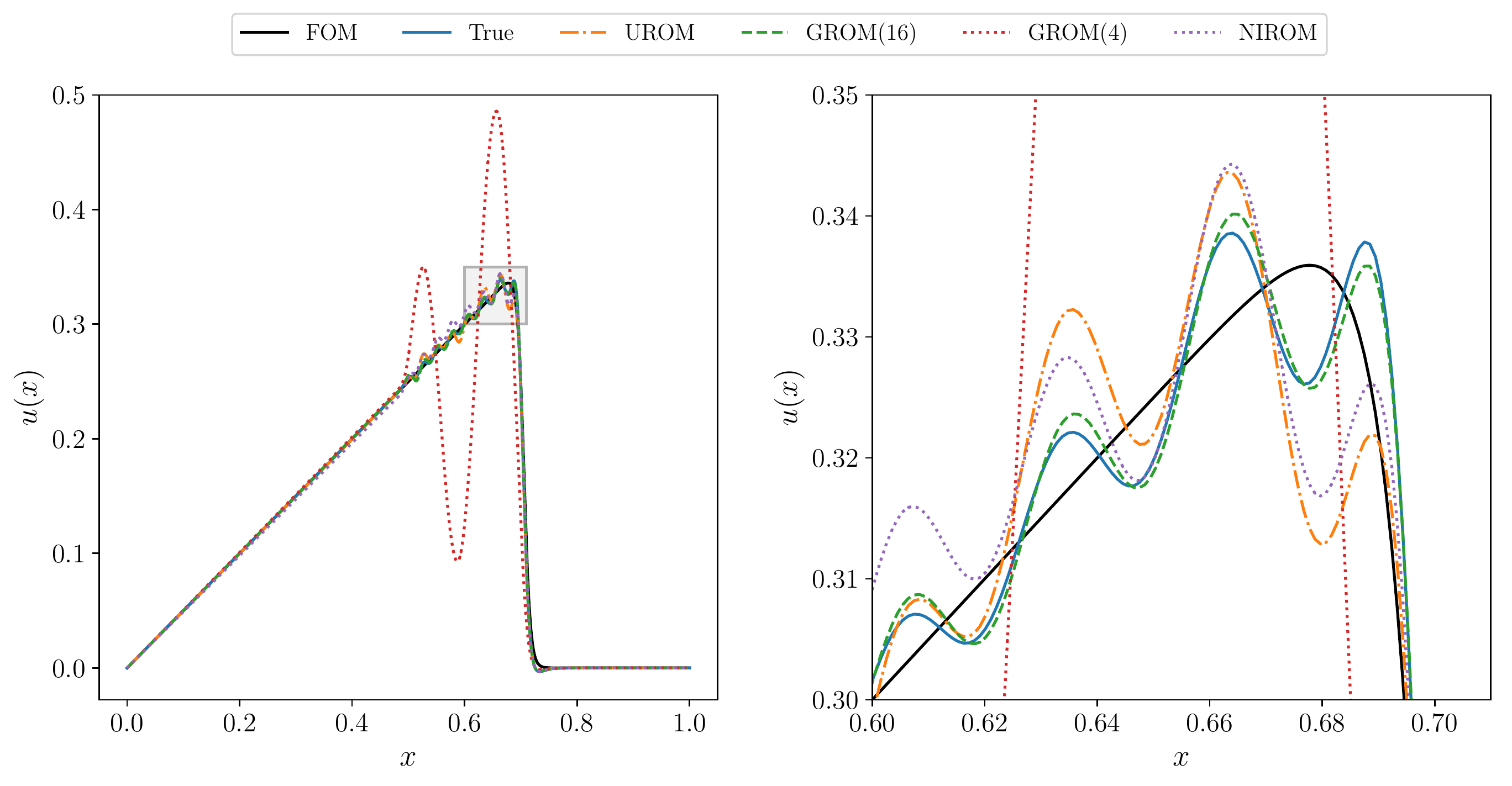}
	\caption{Final velocity fields (at $t=1$) for Burgers problem at $\text{Re} =1000$ with a zoom-in view at the right using $R=4$, and $Q=16$. Oscillations in UROM and GROM($16$) occur mainly because a subspace spanning the first $16$ modes is insufficient to capture the dynamics at this Reynolds number.}
	\label{fig:ufinal1000}
\end{figure}

The temporal field evolution of flow field is shown in Figure~\ref{fig:u1000}, which illustrates the non-physical and unstable behavior of both GROM($4$) and NIROM approaches. The final field is plotted in Figure~\ref{fig:ufinal1000} with a close-up view at the right. It can be seen that even the true projected fields do not match the FOM and show some fluctuations at the shock region. For this type of behavior, a larger subspace is required to capture most of the dynamics of the flow at $\text{Re} =1000$. Alternatively, a principal interval decomposition approach can be adopted to localize POD basis and tailor more representative compact subspaces \cite{ahmed2019memory,dihlmann2011model}.

\subsection{2D Vortex Merger Problem}
As an application for 2D Navier-Stokes equations, we examine the vortex merger problem (i.e., the merging of co-rotating vortex pair) \cite{buntine1989merger}. The merging process occurs when two vortices of the same sign with parallel axes are within a certain critical distance from each other, ending as a single, nearly axisymmetric, final vortex \cite{von2000vortex}. We consider an initial vorticity field of two Gaussian-distributed vortices with a unit circulation as follows,
\begin{align}
    \omega(x, y, 0) &= \exp\left( -\rho \left[ (x-x_1)^2  + (y-y_1)^2 \right] \right) \nonumber  \\
                    &+ \exp{\left( -\rho \left[ (x-x_2)^2 + (y-y_2)^2 \right] \right)},
\end{align}
where $\rho$ is an interacting constant set as $\rho = \pi$ and the vortices centers are initially located at $(x_1,y_1) = \left( \dfrac{3\pi}{4},\pi \right)$ and $(x_2,y_2) = \left( \dfrac{5\pi}{4},\pi \right)$. We use a Cartesian domain $(x,y) \in [0,2\pi] \times [0,2\pi]$ over a spatial grid of size $256^2$, with periodic boundary conditions. For this 2D problem, we collect $200$ snapshots for $t \in [0,20]$, while varying Reynolds number as $\text{Re} \in \{200,400,600,800\}$. Similar to the 1D Burgers problem, we test our framework at $\text{Re} =500$ and $\text{Re} =1000$. Also, for basis interpolations, we used reference points at $\text{Re} =600$ and $\text{Re} =800$, respectively.

\subsubsection{\textbf{Re = 500}: demonstrating interpolatory capability}

\begin{figure}[!ht]
	\centering
	\includegraphics[width=0.95\textwidth]{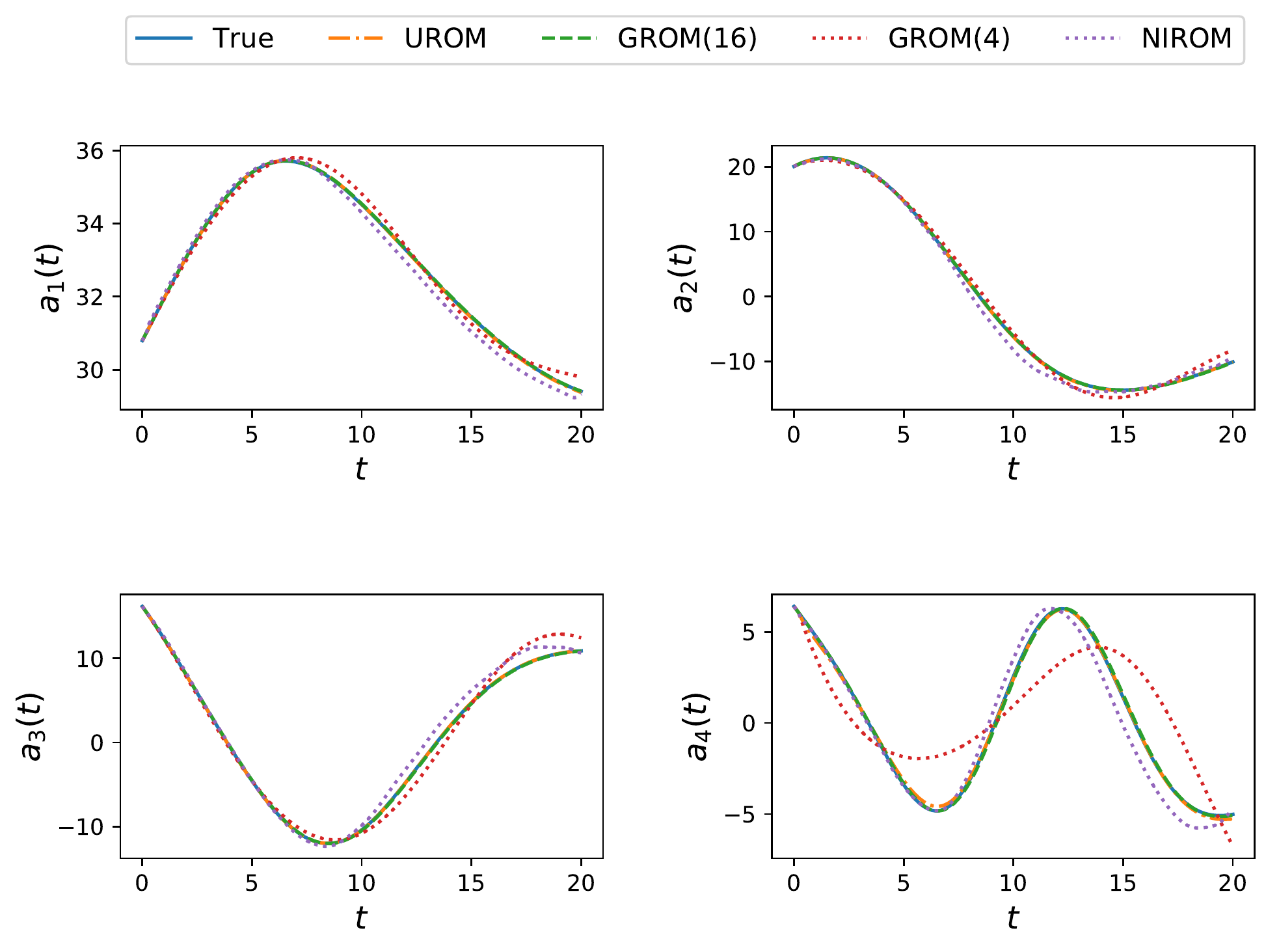}
	\caption{Temporal evolution of the first 4 POD modal coefficients of vorticity field for 2D vortex merger problem as predicted by UROM, GROM(4), GROM(16), and NIROM compared with the true values obtained by projection of FOM field on the interpolated modes at $\text{Re} =500$.}
	\label{fig:vort_a500}
\end{figure}

\begin{figure}[!ht]
	\centering
	\includegraphics[width=0.95\textwidth]{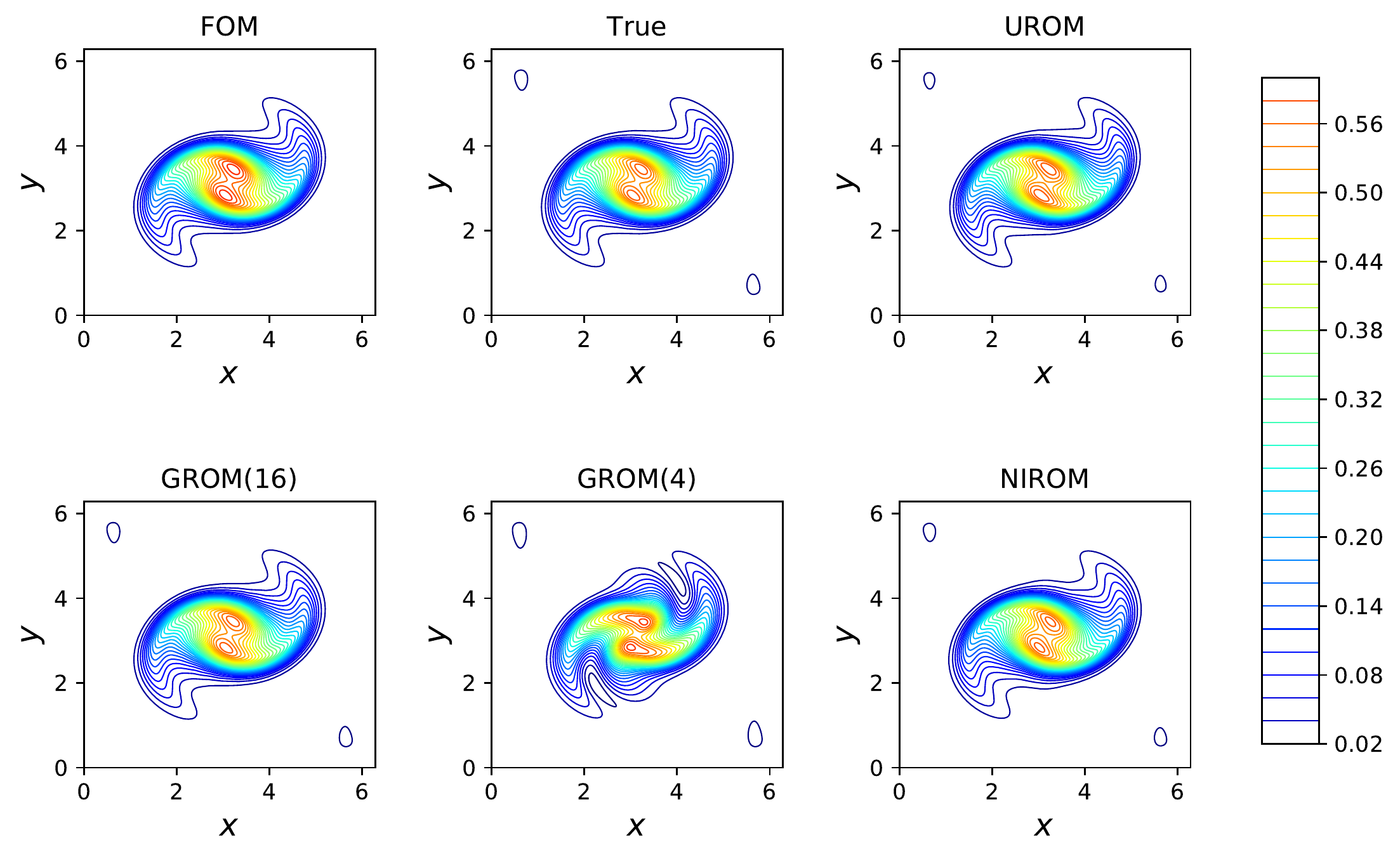}
	\caption{Final vorticity fields (at $t=20$) for 2D vortex merger problem at $\text{Re} =500$ with $R=4$, and $Q=16$.}
	\label{fig:wfinal500}
\end{figure}

Figure~\ref{fig:vort_a500} shows the temporal evolution of the first 4 POD coefficients for the vorticity field. Recall that the temporal coefficients for vorticity and streamfunction fields are the same, as discussed in Section~\ref{sec:NS}. We can see that both UROM and GROM($16$) accurately predict the true modal dynamics. For better visualizations, the final vorticity field at $t=20$ is given in Figure~\ref{fig:wfinal500}, where GROM($4$) is showing instabilities manifested in the reconstructed field. For this interpolatory case, NIROM is providing ``acceptable'' results.

\subsubsection{\textbf{Re = 1000}: demonstrating extrapolatory capability}

\begin{figure}[!ht]
	\centering
	\includegraphics[width=0.95\textwidth]{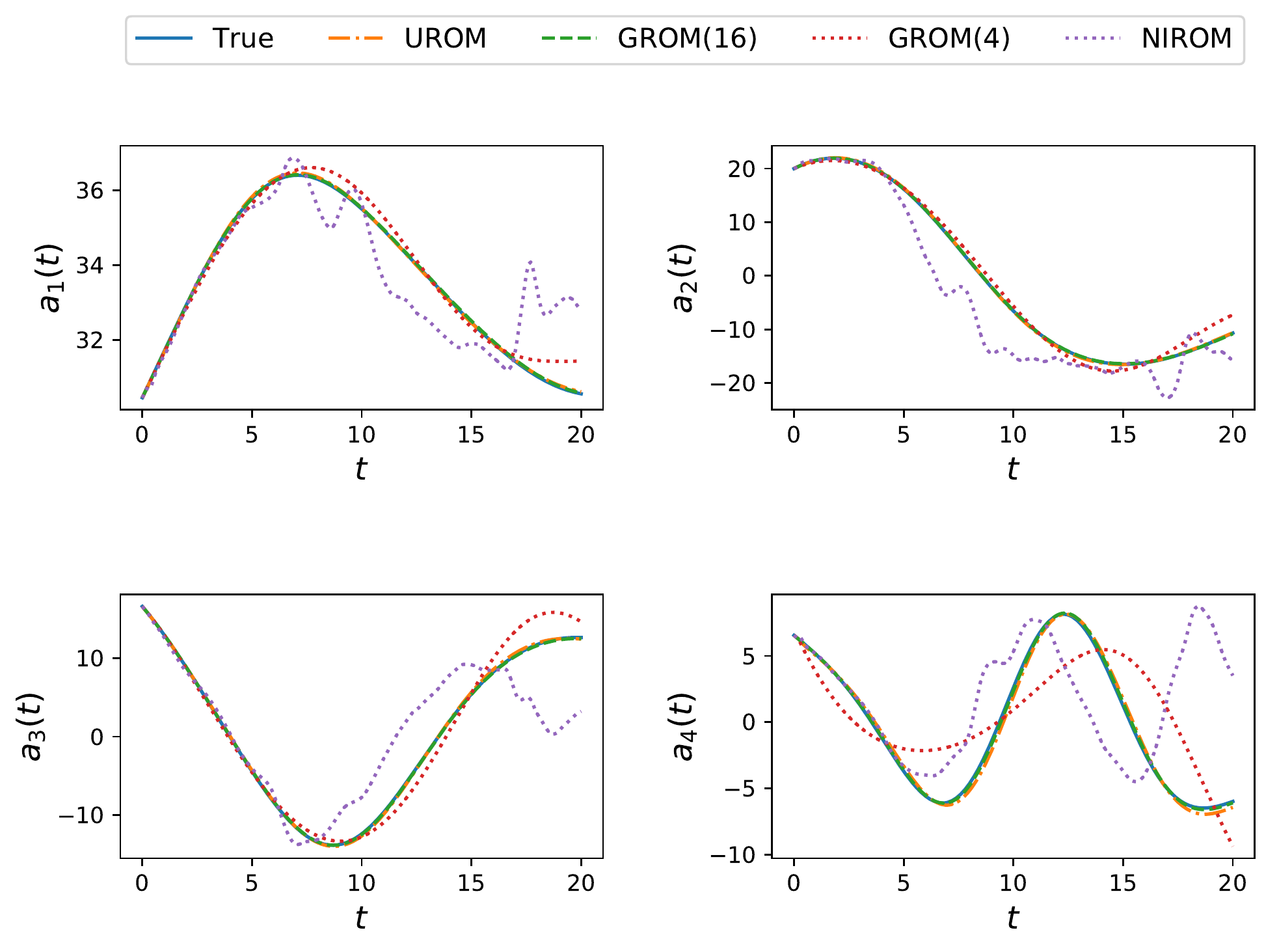}
	\caption{Temporal evolution of the first 4 POD modal coefficients of vorticity field for 2D vortex merger problem as predicted by UROM, GROM(4), GROM(16), and NIROM compared with the true values obtained by projection of FOM field on the interpolated modes at $\text{Re} =1000$.}
	\label{fig:vort_a1000}
\end{figure}

\begin{figure}[!ht]
	\centering
	\includegraphics[width=0.95\textwidth]{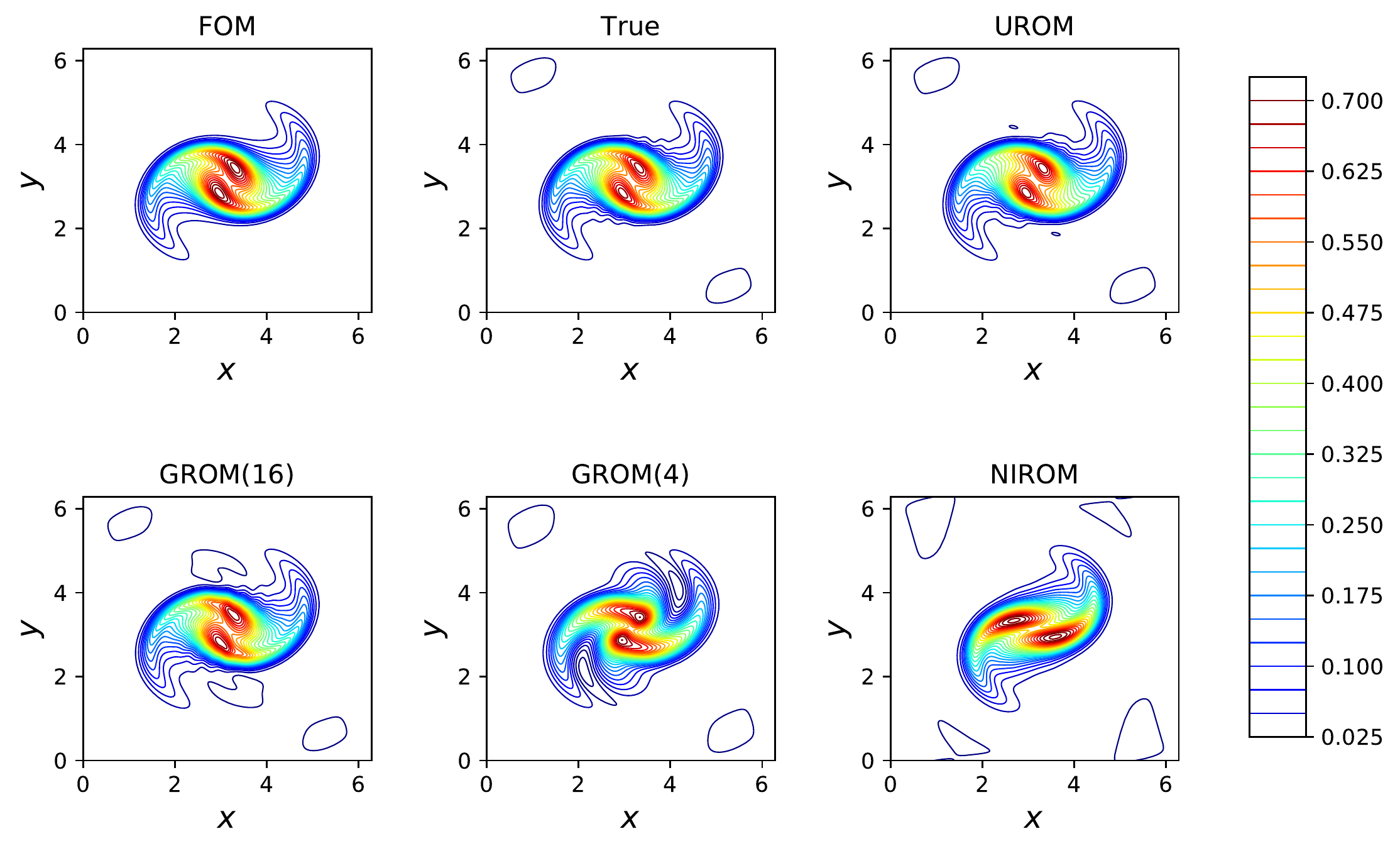}
	\caption{Final vorticity fields (at $t=20$) for 2D vortex merger problem at $\text{Re} =1000$ with $R=4$, and $Q=16$.}
	\label{fig:wfinal1000}
\end{figure}

The investigated approaches, namely GROM, UROM and NIROM, are tested at $\text{Re} =1000$ as a case that is out-of-range compared to the training set. The POD modal coefficients predicted by these approaches are given in Figure~\ref{fig:vort_a1000}. It can be easily seen that as time increases, the predictions of GROM($4$) and NIROM become poor. Using a neural network as time-stepping integrator in NIROM increases its sensitivity to computational noise and this recursive deployment accumulates the error until predictions totally depart from the true trajectory. This is even clearer in the reconstructed field shown in Figure~\ref{fig:wfinal1000}, where the orientation of the merging vortices is not matching the true orientation. GROM($4$) prediction is also suffering from severe deformation of the true flow topology.  On the other hand, the field reconstruction via UROM is accurate compared to the true projection and GROM($16$).

\subsection{Computing Time} \label{sec:cpu}
\begin{figure}[!ht]
	\centering
	\includegraphics[width=0.95\textwidth]{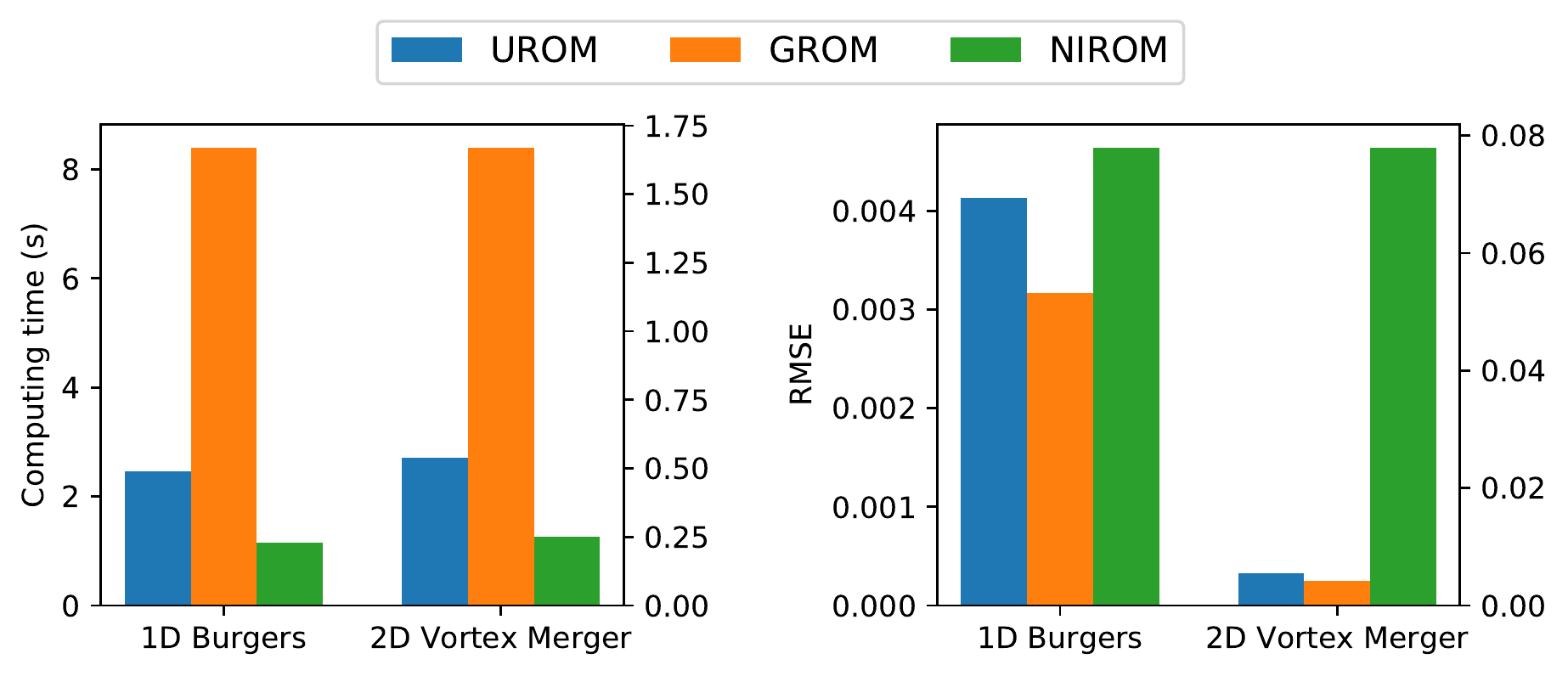}
	\caption{Computing time of testing (online) stage at $\text{Re}=1000$ (left) and RMSE of reconstructed fields at final time for UROM ($4+12$) (i.e., $R=4$ and $Q=16$), GROM($16$), and NIROM($16$).}
	\label{fig:cpu}
\end{figure}

Finally, we report the ``online'' computing time for the investigated approaches. In particular, we show the computational time as well as the root mean squared error (RMSE) of reconstructed fields at final time for the two test cases at $\text{Re} =1000$ in Table~\ref{table:cpu}. The reported RMSE is computed as
\begin{equation}
    \text{RMSE}(t) = \sqrt{\dfrac{1}{N} \sum_{i=1}^{N} \big(\mathbf{u}^{FOM}(\mathbf{x},t) - \mathbf{\hat{u}}(\mathbf{x},t) \big)^2 },
\end{equation}
where $N$ represents the spatial resolution (i.e., $N=N_x\times N_y$). In this table, we also report the NIROM results using 4 modes in the input and output layers. Although GROM($4$) is the fastest, its predictions are very poor and further corrections and stabilization might be required. Also, a subspace spanned by only the first four POD modes might be insufficient in complex applications. On the other hand, GROM($16$) is the slowest. We can also observe that computation time of UROM is close to that of NIROM and much lower than GROM($16$). In Figure~\ref{fig:cpu}, we present a bar chart for both the computing time and RMSE of reconstructed fields at final time to illustrate the time-accuracy trade-off. 

We note that Table~\ref{table:cpu} and Figure~\ref{fig:cpu} document the performance of our implementation rather than that of the approaches. We should emphasize that, in this paper, we are not aiming at benchmarking the computational performance of these approaches. Instead, our main objective is to demonstrate the feasibility of hybrid approaches fusing physics-based and machine learning models. Nonetheless, the runtimes in Table~\ref{table:cpu} indicate that GROM approximately (not exactly due to various other loading/writing abstractions in our Python implementations) scales with $R^3$. Therefore, combining NIROM and GROM, UROM yields better computational performance. We also note that, if written more optimally, we would also expect that the execution time of UROM (with 16 modes) can be reduced to the sum of the execution times of GROM (with 4 modes) and NIROM (with 16 modes). Indeed, we remark that the second LSTM in UROM (representing the map $g$) need not be used at all times and can be deployed only at the instant of interest. In that case, the UROM computing times become $1.31$ s for Burgers case and $0.30$ s for vortex merger (whic are very close to NIROM computing time).


\begin{table}[htbp]
	\centering
	\caption{Computing time (in seconds) and RMSE of UROM, GROM($4$), GROM($16$), NIROM($4$), and NIROM($16$) for $\text{Re}=1000$. We note that the computing time assessments documented in this table are based on Python executions.}
	\begin{tabular}{p{0.2\textwidth}p{0.12\textwidth}p{0.12\textwidth}p{0.12\textwidth}p{0.12\textwidth}} 
		\hline\noalign{\smallskip} 
		Framework & \multicolumn{2}{c}{1D Burgers} &  \multicolumn{2}{c}{2D vortex-merger} \\ 
		 & \ time(s) & RMSE & \quad time(s) & RMSE   \\ 
		\noalign{\smallskip}\hline\noalign{\smallskip}
		UROM        & \ $2.46$ &  $4.13\text{E}-3$ &  \quad $0.54$  & \ $5.44\text{E}-3 $ \\
        GROM($16$)  & \ $8.40$ &  $3.17\text{E}-3$ &  \quad $1.67$  & \ $4.17\text{E}-3 $ \\
        GROM($4$)   & \ $0.17$ &  $5.17\text{E}-2$ &  \quad $0.06$  & \ $3.99\text{E}-2 $ \\
        NIROM($16$) &\ $1.16$  &  $4.64\text{E}-3$ &  \quad $0.25$  & \ $7.80\text{E}-2 $ \\
        NIROM($4$)  &\ $1.07$  &  $3.14\text{E}-2$ &  \quad $0.23$  & \ $8.58\text{E}-2 $ \\
		\hline
	\end{tabular}
	\label{table:cpu}
\end{table}
\section{Concluding Remarks} \label{sec:Conc}
In the present study, we have proposed an uplifted reduced order modeling (UROM) approach to elevate the standard Galerkin projection reduced order modeling (GROM). This approach can be considered as a hybrid approach between physics-based and purely data-driven techniques. With GROM at the core of the framework, UROM (with three modeling layers) enhances the model generalizability and interpretability. Moreover, large scales (represented by the first few modes) are given due attention since they control most of the bulk mass, momentum, and energy transfers. Therefore, two out of a total of three layers in UROM aim at predicting the dynamics of these modes as accurately as possible. Then, an uplifting layer is designed to enhance the prediction resolution (i.e., super-resolution). Performance of UROM has been compared against standard GROM and fully non-intrusive ROM (NIROM) approaches. 

Two test cases, representing convection-dominated flows in 1D and 2D settings, have been used to evaluate the UROM. For testing, two out-of-sample control parameters have been used to study the interpolatory and extrapolatory performances. In all cases, UROM showed very good results, compared to GROM and NIROM. In particular, UROM($4+12$) has been shown to provide more accurate results than both GROM($4$) and NIROM. In contrast to NIROM where the deployment is fully data-driven, the LSTMs in UROM take their inputs from a physics-based approach. This can be considered as one way of leveraging physical information and intuition into LSTM. On the other hand, UROM has provided significant speed-ups compared to GROM($16$) with comparable accuracy. Although we have presented the results for $Q=16$, more complex flows can require much larger $Q$, which makes GROM($Q$) unfeasible. Finally, this UROM approach is thought to open new avenues to utilize data-driven tools to enhance existing physical models as well as use physics to inform data-driven approaches to maximize the pros of both approaches and mitigate their cons.

\section*{Acknowledgements}
This material is based upon work supported by the U.S. Department of Energy, Office of Science, Office of Advanced Scientific Computing Research under Award Number DE-SC0019290. O.S. gratefully acknowledges their support. 

Disclaimer. This report was prepared as an account of work sponsored by an agency of the United States Government. Neither the United States Government nor any agency thereof, nor any of their employees, makes any warranty, express or implied, or assumes any legal liability or responsibility for the accuracy, completeness, or usefulness of any information, apparatus, product, or process disclosed, or represents that its use would not infringe privately owned rights. Reference herein to any specific commercial product, process, or service by trade name, trademark, manufacturer, or otherwise does not necessarily constitute or imply its endorsement, recommendation, or favoring by the United States Government or any agency thereof. The views and opinions of authors expressed herein do not necessarily state or reflect those of the United States Government or any agency thereof.

\bibliographystyle{unsrt} 
\bibliography{ref}   

\end{document}